\documentclass[a4paper,11pt]{article}
\pdfoutput=0 

\usepackage{jheppub} 

\usepackage[T1]{fontenc} 
\usepackage[utf8]{inputenc}
\usepackage{amsfonts}
\usepackage{amsthm}
\usepackage[]{algorithm2e}
\usepackage{mathtools}
\usepackage{axodraw4j}

\title{Transforming differential equations of multi-loop Feynman 
      integrals into canonical form}

\author{Christoph Meyer}
\affiliation{Institut f\"ur Physik, Humboldt-Universit\"at zu Berlin, 12489 Berlin, Germany}

\emailAdd{christoph.meyer@physik.hu-berlin.de}

\abstract{The method of differential equations has been proven to be a powerful tool for the computation of multi-loop Feynman integrals appearing in quantum field theory. It has been observed that in many instances a canonical basis can be chosen, which drastically simplifies the solution of the differential equation. In this paper, an algorithm is presented that computes the transformation to a canonical basis, starting from some basis that is, for instance, obtained by the usual integration-by-parts reduction techniques. The algorithm requires the existence of a rational transformation to a canonical basis, but is otherwise completely agnostic about the differential equation. In particular, it is applicable to problems involving multiple scales and allows for a rational dependence on the dimensional regulator. It is demonstrated that the algorithm is suitable for current multi-loop calculations by presenting its successful application to a number of non-trivial examples.}

\preprint{HU-EP-16/24}

\SetKwRepeat{Do}{do}{end}

\def\eq#1{eq.~\eqref{#1}}

\newcommand*{\ima}{\textup i} 
\newcommand*{\defeq}{=} 

\newcommand*{\varset}{\epsilon, \{x_j\}} 
\newcommand*{\invariants}{\{x_j\}} 

\newtheorem{defn}{Definition}
\newtheorem{thm}{Theorem}
\newtheorem{corollary}{Corollary}
\newtheorem{lemma}{Lemma}

\begin{document}
\maketitle
\flushbottom

\section{Introduction}
With the observation of the Higgs boson in 2012 \cite{ Aad:2012tfa, Chatrchyan:2012xdj} the last missing building block predicted by the standard model has been discovered. Despite this great success of the standard model as fundamental theory of particle physics, it is well known that new physics beyond the standard model exists, neither dark matter nor neutrino masses are explained by the standard model in its present form. 

With a steadily increasing experimental precision, the LHC experiments are currently searching for possible standard model extensions. In the experimental analysis, both direct searches and precision measurements are utilized to look for deviations from the standard model. While the former profit from the increased center of mass energy of run II, the latter benefit from higher statistics and a better understanding of systematic uncertainties. However, in both cases precise theoretical predictions for the signal reactions as well as the background reactions are mandatory.

Since most processes at the LHC are dominated by QCD, leading-order predictions suffer in general from large theoretical uncertainties and provide only a rough estimate of the respective cross sections. Higher order corrections in the perturbative expansion are therefore necessary in order to achieve percent level precision. 

As far as next-to-leading order (NLO) calculations are considered, tremendous progress has been made in the last twenty years. Today, NLO calculations are often considered as an algorithmically solved problem. Various publicly available tools (cf. \cite{Ossola:2015xga} and references therein) allow the automated calculation of NLO corrections for typical LHC processes. However, as the recent calculation of higher order corrections for Higgs production via gluon fusion \cite{Anastasiou:2015ema} illustrates, in general NLO is not sufficient to achieve theoretical accuracies below ten percent and even higher order corrections are required.

Beyond NLO, the same level of maturity has not yet been reached. While for some of the steps required in multi-loop calculations at least in principle solutions exist --- the practical application is often limited by the available computer resources --- , the evaluation of the scalar multi-loop integrals still represents a major bottleneck. In particular, no general algorithm is known to perform this integration in an automated way. This situation is very different from the NLO case where all the scalar one-loop integrals are known. It was in fact this knowledge combined with field theoretical insights which triggered the progress in NLO calculations mentioned above. A better understanding of multi-loop integrals is thus crucial for any progress beyond the one-loop level.
In higher order calculations various techniques have been employed to evaluate the required multi-loop integrals. One of the most powerful ones is the method of differential equations \cite{Kotikov:1990kg, Remiddi:1997ny, Gehrmann:1999as}.

A major improvement of the differential equations approach has been made by the observation that very often a particularly simple form of the differential equation can be achieved by changing the basis of integrals  \cite{Henn:2013pwa}. In this form, often called \emph{canonical} or \emph{$\epsilon$-form}, the integration is --- up to the determination of integration constants --- reduced to a merely combinatorial task. This refined method has been successfully applied to many recent multi-loop calculations \cite{Henn:2013pwa, Henn:2013fah, Henn:2013woa, Henn:2013nsa, Argeri:2014qva, Henn:2014lfa, Caron-Huot:2014lda, Gehrmann:2014bfa, Caola:2014lpa, Li:2014bfa, Hoschele:2014qsa, DiVita:2014pza, vonManteuffel:2014mva, Grozin:2014hna, Bell:2014zya, Huber:2015bva, Gehrmann:2015ora, Gehrmann:2015dua, Bonciani:2015eua, Anzai:2015wma, Grozin:2015kna, Gehrmann:2015bfy, Gituliar:2015iyq, Lee:2016htz, Henn:2016men, Bonciani:2016ypc, Eden:2016dir, Lee:2016lvq, Bonciani:2016qxi, Bonetti:2016brm}. Often, the most difficult part of these calculations is to find a basis of master integrals in which the differential equation attains a canonical form. 
Several methods to find a canonical form have already been discussed \cite{Henn:2013pwa, Argeri:2014qva, Caron-Huot:2014lda, Gehrmann:2014bfa, Hoschele:2014qsa, Lee:2014ioa, Henn:2014qga, Eden:2016dir}. For the algorithm presented in \cite{Lee:2014ioa} there is an implementation publicly available \cite{Gituliar:2016vfa}. However, this method is limited to the case of integrals depending on only one dimensionless scale. The aim of this article is to present an algorithm allowing to compute a canonical basis --- provided it exists --- in the most general case of differential equations depending on multiple scales with a rational dependence on the dimensional regulator. The proposed algorithm has been implemented in \texttt{Mathematica} and has been successfully tested on non-trivial examples \cite{Meyer:0000abc}. The outline of this article is as follows. In section \ref{sec:Preliminaries}, basic properties of the differential equations are recalled, which also serves the purpose of fixing the notation. Based on the assumption that a rational transformation exists that transforms a differential equation into canonical form, section \ref{sec:Algorithm} first explores some general features of such transformations, which are useful for devising the algorithm. Subsequently, it is shown that the transformation may be obtained as a rational solution of a finite number of differential equations. Using a generalized partial fractions technique \cite{Lei78, 2012arXiv1206.4740R}, it is argued that rational solutions of these equations can be expressed as a linear combinations of a particular class of rational functions. Section \ref{sec:Applications} discusses the application of the presented algorithm to double box topologies, which are relevant for NNLO corrections to single top-quark production \cite{Brucherseifer:2014ama, Assadsolimani:2014oga} and vector boson pair production \cite{Cascioli:2014yka, Gehrmann:2014fva, Grazzini:2016swo}. The full results of these examples are provided in ancillary files. The conclusions are drawn in section \ref{sec:Conclusion}. For easy reference, appendix \ref{App:PolyAlgebra} lays out standard definitions and results about polynomial algebra that are needed in section \ref{sec:Algorithm}.

\section{Preliminaries}
\label{sec:Preliminaries}
Higher order corrections in quantum field theory involve integrations
over the unconstrained loop momenta, in general in the form of tensor
integrals. It is straightforward to express the tensor integrals in terms of scalar
integrals with raised powers of the propagators \cite{Passarino:1978jh, Davydychev:1991va, Tarasov:1996br, Tarasov:1997kx, Anastasiou:1999bn, Glover:2004si}:
\begin{equation}
I(\nu_1,\dots,\nu_n)=\int\prod_{i=1}^L\frac{\textup{d}^dl_i}{\ima\pi^{d/2}}\frac{1}{P^{\nu_1}_1\cdots P_n^{\nu_n}},
\end{equation}
where the $P_i$ denote inverse propagators, which are functions of the loop momenta, the external momenta and the masses of the particles running in the loops. For a given set of propagators, each integral is assigned a \emph{sector-id} by 
\begin{eqnarray}
ID[I]&=&\sum_{k=1}^n2^{k-1}\Theta(\nu_k),\\
\Theta(x)&=&\begin{cases} 
      1 & x > 0 \\
      0 & x\leq 0.
   \end{cases}
\end{eqnarray}
A set of integrals with the same sector-id is called a \emph{sector}. Note that a sector is completely specified by the set of propagators with positive powers. A sector is said to be a \emph{subsector} of another sector if its set of propagators with positive powers is a proper subset of the other sectors set. 

In practice, the tensor reduction often leads to a large number of scalar integrals. However, there also exists a large number of linear relations among them. These integration-by-parts identities \cite{Tkachov:1981wb, Chetyrkin:1981qh, Gluza:2010ws, Schabinger:2011dz, Kant:2013vta, vonManteuffel:2014ixa, Larsen:2015ped} and Lorentz identities  \cite{Gehrmann:1999as} can be used to express all scalar integrals as linear combinations of a finite number of independent master integrals \cite{Smirnov:2010hn, Lee:2013hzt}. The reduction to master integrals --- while in practice still challenging --- is often considered as a solved problem. For the Laporta algorithm \cite{Laporta:2001dd}, which allows to systematically perform this reduction, there are various implementations publicly available \cite{Anastasiou:2004vj, Studerus:2009ye, vonManteuffel:2012np, Smirnov:2013dia, Smirnov:2014hma}. A different strategy has been presented and implemented in \cite{Lee:2012cn}.

As mentioned before, one method to attempt the evaluation of the master integrals is to derive a set of coupled differential equations for the $m$-dimensional vector of master integrals $\vec{f}(\varset)$ \cite{Kotikov:1990kg, Remiddi:1997ny, Gehrmann:1999as}. The master integrals are assumed to be normalized such that their mass-dimension is zero. Then, the master integrals can be considered as functions of a set $\invariants$ of $M$ dimensionless kinematic invariants and the dimensional regulator $\epsilon$ defined through $d=4-2\epsilon$, where $d$ denotes number of spacetime dimensions. The differential equation is obtained by taking the derivatives of $\vec{f}(\varset)$ with respect to all kinematic invariants. Each derivative of a master integral equals a linear combination of scalar integrals with the same or lower sector-id. Applying the Laporta reduction to these integrals, one may express the derivative of a master integral again as a linear combination of master integrals. Thus, upon differentiating with respect to all kinematic invariants, the following linear system of differential equations is obtained
\begin{equation}
\label{DEQNonDifferentialForm}
\partial_i\vec{f}(\varset)=a_{i}(\varset)\vec{f}(\varset),\quad i=1,\dots, M,
\end{equation}
with the $a_i(\varset)$ being $m\times m$ matrices of rational functions in the kinematic invariants $\invariants$ and $\epsilon$. The fact that the matrices $a_i(\varset)$ are rational functions of the kinematic invariants and $\epsilon$ follows from the structure of the integration-by-parts relations. The reduction to master integrals can be done such, that each scalar integral is expressed as a linear combination of master integrals with the same or lower sector-id. If the components of $\vec{f}(\varset)$ are then ordered by their sector-ids, the $a_{i}$ attain a block-triangular form where each sector corresponds to a block. In the more compact differential notation \eq{DEQNonDifferentialForm} can be written as
\begin{equation}
\label{DEQDifferentialForm}
\textup{d}\vec{f}(\varset)=a(\varset)\vec{f}(\varset),
\end{equation}
with 
\begin{equation}
a(\varset)=\sum_{i=1}^Ma_i(\varset)\textup{d}x_i.
\end{equation}
Taking the exterior derivative of \eq{DEQDifferentialForm} implies
the following integrability condition
\begin{equation}
\label{diffIntegralityCondition}
\textup{d}a-a\wedge a=0,
\end{equation}
where it has been used that the master integrals are
linearly independent over the field of rational functions in the
invariants.  In practice, this condition can serve as a consistency
check of the differential equation.

Transforming the basis of master integrals with an invertible transformation $T$, 
\begin{equation}
\vec{f}=T(\varset)\vec{f}^\prime,
\end{equation}
as suggested in \cite{Henn:2013pwa}, leads to the following transformation law for $a(\varset)$:
\begin{equation}
\label{DEQTrafo}
a^\prime = T^{-1}aT-T^{-1}\textup{d}T.
\end{equation}
The differential equation is said to be in \emph{dlog}-form if $a(\varset)$ can be written as follows
\begin{equation}
\textup{d}A(\varset)=a(\varset),
\end{equation}
with
\begin{equation}
A(\varset)=\sum_{l=1}^NA_l(\epsilon)\log(L_l(\invariants)).
\end{equation}
Here $L_l(\invariants)$ denotes polynomials in the kinematic invariants and the $A_l$ are $m\times m$ matrices, which solely depend on $\epsilon$. The set of polynomials
\begin{equation}
\mathcal{A}=\{L_1(\invariants),\dots,L_N(\invariants)\}
\end{equation}
is commonly referred to as the \textit{alphabet} of the differential equation. The individual polynomials are called the \textit{letters} of the differential equation. 
In \cite{Henn:2013pwa} it was observed that with a suitable change of the basis of master integrals it is often possible to arrive at a form in which the dependence on $\epsilon$ factorizes: 
\begin{equation}
\label{EpsForm}
A(\varset)=\epsilon\sum_{l=1}^N\tilde{A}_l\log(L_l(\invariants)),
\end{equation}
with $\tilde{A}_l$ being constant $m\times m$ matrices.  In this form, which is called \textit{canonical form} or \textit{$\epsilon$-form}, it is particularly easy to solve the differential equation in terms of iterated integrals \cite{Chen:1977oja, Goncharov:1998kja}.

\section{Algorithm}
\label{sec:Algorithm}

Finding a basis in which the differential equation attains $\epsilon$-form is in general a highly
non-trivial task. In fact, it is not even known whether such a basis can always be found for master integrals that evaluate to multiple polylogarithms. It is assumed throughout this section that an $\epsilon$-form exists for the $a(\varset)$ under consideration and that it can be attained with a rational transformation. In addition to that, it is assumed that $a(\varset)$ itself is also rational in $\epsilon$ and the invariants. By making only these mild assumptions, it is possible to both learn more about general properties of the problem and construct an algorithm, which has a broad scope of application.

In subsection \ref{sec:GenProps} some general properties of a transformation to an $\epsilon$-form are presented. Then, it is shown in subsection \ref{sec:ExpTrafo} that such transformations are determined by a finite number of differential equations, which are obtained by expanding a reformulated version of the transformation law. An extension of this strategy to off-diagonal blocks is described in subsection \ref{sec:OffDiagPart}, allowing for a more efficient recursive application of the algorithm. A generalized partial fractions technique is described in subsection \ref{sec:Leinartas}, which is used in subsection \ref{sec:Solving} to solve the aforementioned differential equations for a rational transformation.

\subsection{General properties of the transformation}
\label{sec:GenProps}
It is useful to first look into general properties of the transformation law \eq{DEQTrafo}. Let $T$ be a transformation that transforms the differential equation into $\epsilon$-form. Then an $\tilde{A}$ exists such that
\begin{equation}
\epsilon\textup{d}\tilde{A}(\invariants)=a^\prime(\varset)
\end{equation}
holds. In this case, \eq{DEQTrafo} can be written in the form
\begin{equation}
\label{aprimeINeps}
\epsilon\textup{d}\tilde{A}=T^{-1}aT-T^{-1}\textup{d}T.
\end{equation}
A simple, but important observation is that a subsequent constant transformation $C$
does not spoil the $\epsilon$-form of the differential equation, as it leads to
\begin{equation}
\label{constTpreserves}
a^{\prime\prime}=\epsilon\sum_{l=1}^N\left(C^{-1}\tilde{A}_lC\right)\textup{d}\log(L_l(\invariants)),
\end{equation}
which is again in $\epsilon$-form. Similarly, a subsequent
transformation of the form $T=g(\epsilon)\mathbb{I}$
 with a nonzero rational function $g(\epsilon)$ does not alter the differential equation at all and thus in particular preserves the $\epsilon$-form.

Taking the trace on both sides of \eq{aprimeINeps} leads to
\begin{equation}
\epsilon\textup{Tr}[\textup{d}\tilde{A}]=\textup{Tr}[a]-\textup{Tr}[T^{-1}\textup{d}T].
\end{equation}
Applying Jacobi's formula for the differential of determinants
\begin{equation}
\textup{d}\det(T)=\det(T)\textup{Tr}[T^{-1}\textup{d}T],
\end{equation}
leads to
\begin{equation}
\label{TrOfTrafo}
\textup{d}\log(\det(T))=\textup{Tr}[a]-\epsilon\textup{Tr}[\textup{d}\tilde{A}].
\end{equation}
It follows that a necessary condition for the existence of an $\epsilon$-form is that $\textup{Tr}[a]$ is of the following form:
\begin{equation}
\textup{Tr}[a]=\textup{d}\left(\epsilon\textup{Tr}[\tilde{A}]+\log(\det(T))\right).
\end{equation}
In fact, with \eq{EpsForm} it is evident that $\textup{Tr}[a]$ has to be in dlog-form
\begin{equation}
\label{Tradlog}
\textup{Tr}[a]=\epsilon\sum_{l=1}^N\textup{Tr}[\tilde{A}_l]\textup{d}\log(L_l(\invariants))+\textup{d}\log(\det(T)).
\end{equation}
Note that the term dlog-form is used here in a more general sense, since $\det(T)$ may depend on $\epsilon$. As the components of $T$ are required to be rational in the invariants and $\epsilon$, $\det(T)$ will also have this property. 
Therefore, the summands of $\det(T)$ can be put on a common denominator and the resulting numerator and denominator polynomials can then be factorized into irreducible polynomials in $K[\varset]$. Here, $K[\varset]$ denotes the ring of polynomials in the invariants and $\epsilon$ with coefficients in a field $K$. There is no need to specify the field at this point, for the present application one may have the real or complex numbers in mind. Thus, $\det(T)$ can be written as 
\begin{equation}
\det(T)=F(\epsilon)p_1(\invariants)^{e_1}\cdots p_K(\invariants)^{e_K}q_1(\varset)^{d_1}\cdots q_L(\varset)^{d_L},
\end{equation}
with $e_i\in\mathbb{Z}$ and $d_j\in\mathbb{Z}$. The irreducible factors, which only depend on the invariants, are labeled by $p$ and those, which depend on both $\epsilon$ and the invariants, are labeled by $q$. The product of all factors that solely depend on $\epsilon$ is denoted by $F(\epsilon)$. The factorization allows to rewrite \eq{Tradlog}
\begin{eqnarray}
\label{NecessaryCondition}
\textup{Tr}[a]=\epsilon X(\invariants)+Y(\varset)
\end{eqnarray}
with
\begin{eqnarray}
X(\invariants)&=&\sum_{l=1}^N\textup{Tr}[\tilde{A}_l]\textup{d}\log(L_l(\invariants)),\\
Y(\varset)&=&\sum_{i=1}^Ke_i\textup{d}\log(p_i(\invariants))+\sum_{j=1}^Ld_j\textup{d}\log(q_j(\varset)).\\
\end{eqnarray}
This equation can be understood as a necessary condition on the form of $\textup{Tr}[a]$ for a rational transformation $T$ to exist that transforms the differential equation into $\epsilon$-form. In particular, it implies 
\begin{equation}
\textup{Tr}[a^{(k)}]=0, \quad\forall\,k<0,
\end{equation}
where the $a^{(k)}$ denote the coefficients of the $\epsilon$-expansion of $a(\varset)$. The coefficients of the dlog-terms stemming from $\det(T)$ are integers, whereas the coefficients of the dlog-terms from $\textup{Tr}[\textup{d}\tilde{A}]$ are proportional to $\epsilon$. The determinant of $T$ can therefore be calculated up to a rational function $F(\epsilon)$. Moreover, the traces of the $\tilde{A}_l$ of the resulting $\epsilon$-form can be read of as well. In practice, it can be tested whether $\textup{Tr}[a]$ is of the form \eqref{NecessaryCondition}. If this is not the case, it follows that no rational transformation exists that transforms $a(\varset)$ into $\epsilon$-form. Otherwise, it is possible to extract
\begin{eqnarray}
\label{DetIsFixed}
\det(T)&=&F(\epsilon)\exp\left(\int_\gamma Y(\varset)\right),\\
\label{TraceIsFixed}
\textup{Tr}[\textup{d}\tilde{A}]&=&X(\invariants),
\end{eqnarray}
from the coefficients of the dlog-terms. As will be argued later, both equations provide useful information for the determination of $T$. Often, the factors $q_j$ are absent and therefore
$Y(\varset)=Y(\invariants)$. In this case, the above observations turn into
statements about the coefficients of the $\epsilon$-ex\-pan\-sion of
$a(\varset)$:
\begin{eqnarray}
\det(T)&=&F(\epsilon)\exp\left(\int_\gamma \textup{Tr}[a^{(0)}]\right),\\
\textup{Tr}[\textup{d}\tilde{A}]&=&\textup{Tr}[a^{(1)}].
\end{eqnarray}
Furthermore, \eq{NecessaryCondition} implies in this case
\begin{equation}
\textup{Tr}[a^{(k)}]=0, \quad\forall\,\, k\neq 0, 1.
\end{equation}
Note that for one-dimensional sectors \eq{DetIsFixed} already fixes
the transformation up to a rational function in $\epsilon$. It was shown earlier that the choice of this function does not alter the resulting
$a^\prime$. Therefore, the undetermined $F(\epsilon)$ may be set
\begin{equation}
F(\epsilon)=1,
\end{equation}
which then completely fixes the transformation. The determinant provides valuable information for the computation of $T$ for higher-dimensional sectors as well.

\subsection{Expanding the transformation}
\label{sec:ExpTrafo}
Every invertible transformation $T$ that transforms the differential equation into $\epsilon$-form has to satisfy \eq{aprimeINeps} for some $\textup{d}\tilde{A}$, which has to be determined as well. For invertible $T$, \eq{aprimeINeps} can equivalently be written as
\begin{equation}
\label{DEQTrafoAlternativ}
\textup{d}T-aT+\epsilon T\textup{d}\tilde{A}=0.
\end{equation}
This form has the advantage of not containing the inverse of $T$. The strategy to find a solution of this equation is to expand $T$ in $\epsilon$ and solve for its coefficients order by order. 

\subsubsection*{Reformulation in terms of quantities with finite expansion}

In general, the $\epsilon$-expansion of $T$ may have infinitely many non-vanishing coefficients. This poses a problem for the algorithmic computation of these coefficients. In the following, it will be shown how this problem can be circumvented. 

It is evident that \eq{DEQTrafoAlternativ} is invariant under the multiplication of $T$ by a rational function $g(\epsilon)$. Any such rational function can be written as a product of some power of $\epsilon$ and a rational function $\eta(\epsilon)$ with non-vanishing constant coefficient
\begin{equation}
g(\epsilon)=\epsilon^\tau\eta(\epsilon).
\end{equation}
One part of the freedom to chose $g(\epsilon)$ can be exploited by demanding the expansion of $T$ to start at order $\epsilon^0$
\begin{equation}
\label{TStartsAtZero}
T=\sum_{n=0}^\infty\epsilon^nT^{(n)}, \quad T^{(0)}\neq 0.
\end{equation}
This condition only fixes the value of $\tau$ and leaves $\eta(\epsilon)$ unaffected.

As $a(\varset)$ is required to be rational in both the invariants and $\epsilon$, a polynomial $h(\varset)$ exists such that $\hat{a}\defeq ah$ has a finite Taylor expansion in $\epsilon$
\begin{equation}
\label{aStartsAtZero}
\hat{a}=\sum_{k=0}^{k_\textup{max}}\epsilon^k\hat{a}^{(k)}.
\end{equation}
Likewise, there exists a polynomial $f(\varset)$ such that $\tilde{T}\defeq Tf$ has a finite expansion in $\epsilon$
\begin{equation}
\label{TtildeStartsAtZero}
\tilde{T}=\sum_{q=0}^{q_\textup{max}}\epsilon^q\tilde{T}^{(q)},\quad \tilde{T}^{(0)}\neq 0.
\end{equation}
Note that eqs. \eqref{TStartsAtZero} and \eqref{TtildeStartsAtZero} imply that the expansion of $f$ starts at the constant term
\begin{equation}
\label{fStartsAtZero}
f(\varset)=f^{(0)}(\invariants)+\mathcal{O}(\epsilon), \quad f^{(0)}\neq 0,
\end{equation}
whereas \eq{aStartsAtZero} may require the expansion of $h$ to start at some higher order $l_\textup{min}$
\begin{equation}
h(\varset)=\sum_{l=l_\textup{min}}^{l_\textup{max}}\epsilon^l h^{(l)}(\invariants),\quad l_\textup{min}\geq 0.
\end{equation}
This stems from the fact that $a(\varset)$ can in general have negative powers of $\epsilon$ in its expansion, which in the case of $T$ have already been absorbed by the choice of $\tau$. In addition to the above conditions, $h$ and $f$ are required to be \textit{minimal} in the sense that they shall have the smallest possible number of irreducible factors for which $\hat{a}$ and $\tilde{T}$ have finite $\epsilon$-expansions of the above form. This fixes $h$ and $f$ up to multiplicative constants, which are irrelevant here. Let the factorizations of $h$ and $f$ into irreducible factors in $K[\varset]$ be denoted by
\begin{equation}
f=\prod_{i=1}^{N_f}f_i,\quad h=\prod_{i=1}^{N_h}h_i.
\end{equation}
\subsubsection*{Investigating the relation of $f$ and $h$}

It is straightforward to compute $h$ for a given $a(\varset)$. However, $f$ could only be computed directly if $T$ was known. Since this is not the case, the relation of $f$ and $h$ will be investigated in the following. With the above definitions \eq{DEQTrafoAlternativ} reads
\begin{equation}
\label{DEQTrafof}
\frac{h\tilde{T}\textup{d}f}{f}=\left(\textup{d}\tilde{T}+\epsilon\tilde{T}\textup{d}\tilde{A}\right)h-\hat{a}\tilde{T}.
\end{equation}
The right-hand side of \eq{DEQTrafof} only consists of sums and products of quantities, which are assumed to have a finite expansion. Therefore, both sides of the above equation have a finite expansion. For the left-hand side this means that
\begin{equation}
\label{FiniteSummands}
\frac{h\tilde{T}\textup{d}f}{f}=\sum_{i=1}^{N_f}\frac{h\tilde{T}\textup{d}f_i}{f_i}
\end{equation}
has a finite expansion. In the following, it is shown that already each summand of the above sum has a finite expansion. Note that it is sufficient to show that there is no number $n$ of such terms with infinite expansion that can sum up to give a finite expansion. For $n=1$ this is obvious and therefore it remains to be shown that if the assertion holds for $n$ terms, it also holds for $n+1$ terms. Consider $f_1, \dots, f_{n+1}$ and assume that the assertion is not true, i.e. each $h\tilde{T}\textup{d}f_i/f_i$ has an infinite expansion but the sum of all of these terms has a finite expansion. Defining $F_n=f_1\cdots f_n$ one may write
\begin{equation}
\label{Ffsum}
\frac{h\tilde{T}\textup{d}F_n}{F_n}+\frac{h\tilde{T}\textup{d}f_{n+1}}{f_{n+1}}=\frac{h\tilde{T}\textup{d}(F_nf_{n+1})}{F_nf_{n+1}}.
\end{equation}
The second term on the left-hand side has by assumption an infinite expansion and the first term has to have an infinite expansion because the assertion holds for $n$ terms. Since the right-hand side is assumed to have a finite expansion, both $F_n$ and $f_{n+1}$ have to be canceled by corresponding factors in the numerator. However, neither $h$ nor $\tilde{T}$ can be a product of one or both of these factors with a quantity with finite expansion, since this would render the expansions of the terms on the left-hand side finite. Thus, the only possibility left to investigate is 
\begin{equation}
\label{dfargument}
\textup{d}(F_nf_{n+1})=r(\varset)F_nf_{n+1},
\end{equation}
here $r(\varset)$ denotes a rational form with finite expansion. Upon integration, this relation leads to 
\begin{equation}
F_nf_{n+1}=\rho(\epsilon)\exp\left(\int_\gamma r(\varset)\right),
\end{equation}
with $\rho$ denoting a polynomial in $\epsilon$. Since $F_n\cdot f_{n+1}$ is polynomial in the invariants and $\epsilon$, the finiteness of the expansion of $r$ implies
\begin{eqnarray}
r^{(k)}(\invariants)&=&0,\quad\forall\,\,k\neq 0,\\
r^{(0)}(\invariants)&=&\textup{d}\log\left(p(\invariants)\right),
\end{eqnarray}
with $p$ being a polynomial in the invariants. However, since $f$ is required to be minimal, it cannot contain any irreducible factors that are independent of $\epsilon$ and therefore $p$ has to be a constant and thus one finds
\begin{equation}
\label{Fnfnp1epspoly}
F_nf_{n+1}=\rho(\epsilon).
\end{equation}
Both, $F_n$ and $f_{n+1}$ need to have non-vanishing differentials, because otherwise both terms on the left-hand side of \eq{Ffsum} would have a finite expansion. Consequently, both factors have a non-trivial dependence on the invariants. Since $F_n$ and $f_{n+1}$ are polynomials, their product has a non-trivial dependence on the invariants as well, which contradicts \eq{Fnfnp1epspoly}. Thus, the assertion has to be true for $n+1$ terms as well and therefore, by induction, hold for all $n>0$. Altogether, this shows that each summand in \eq{FiniteSummands} has to have a finite expansion.

The minimality of $f$ implies that $\tilde{T}$ cannot be of the form $\tilde{T}=rf_i$ for some rational $r(\varset)$ with finite expansion, because otherwise the factor $f_i$ would not be necessary to render the expansion of $T$ finite and consequently $f$ would not be minimal. Also note that the minimality of $f$ implies that its irreducible factors must 
\emph{all} depend non-trivially on both $\epsilon$ and the invariants. There are only the following two possibilities for a summand of \eqref{FiniteSummands} to have a finite expansion:
\begin{equation}
\label{TwoPossibilities}
\textup{d}f_i=r_if_i\quad \vee\quad h=r_if_i,
\end{equation}
where again $r_i$ denotes a rational function of the invariants and $\epsilon$ that has a finite expansion. However, since the left hand sides of \eqref{TwoPossibilities} are polynomial, a denominator of $r_i$ would have to be canceled by $f_i$, but this would imply that $r_i$ has an infinite expansion. Thus, $r_i$ has in fact to be a polynomial. The first of the above possibilities implies $f_i=c_i(\epsilon)$ by an argument analogous to the one around \eqref{dfargument}, where $c_i(\epsilon)$ denotes an irreducible polynomial in $\epsilon$. In the second case, $f_i$ is equal to one of the irreducible factors of $h(\varset)$. Thus, the irreducible factors of $f$ that are not given by an irreducible factor of $h$ are independent of the invariants.

\subsubsection*{Obtaining a finite expansion with $h$}

As mentioned above, $f$ cannot be used to render the transformation finite, since it cannot be determined directly prior to the computation of $T$. In the following it will be argued how this can be overcome by exploiting the relation of $f$ and $h$ and by using the remaining freedom in the choice of $\eta(\epsilon)$.

Let $\mathcal{S}$ denote the set of indices of the irreducible factors of $h$, which both depend non-trivially on the invariants and are equal to an irreducible factor of $f$. The product of all irreducible factors of $f$ that only depend on $\epsilon$ is denoted by $c(\epsilon)$. Using this notation, $f$ can be written as follows
\begin{equation}
f=c(\epsilon)\prod_{i\in\mathcal{S}\subseteq\{1,\dots,N_h\}}h_i.
\end{equation}
From \eq{fStartsAtZero} it is clear that $c(\epsilon)$ is of the following form
\begin{equation}
c(\epsilon)=c^{(0)}+\mathcal{O}(\epsilon),\quad c^{(0)}\neq 0.
\end{equation}
The remaining freedom in the choice of the overall factor $g(\epsilon)$ can be used to absorb $c(\epsilon)$ by demanding $\eta(\epsilon)=c(\epsilon)$. This completely fixes $g(\epsilon)$ and reduces $f$ to
\begin{equation}
f=\prod_{i\in\mathcal{S}\subseteq\{1,\dots,N_h\}}h_i.
\end{equation}
Although $f$ contains the smallest possible number of irreducible factors that is needed in order to render the expansion of transformation finite, it cannot directly be used in practice, since the set $\mathcal{S}$ is a priori unknown. However, by multiplying with all irreducible factors of $h$, the resulting transformation will also have a finite expansion. This amounts to defining $\hat{T}\defeq Th$, which can now easily be seen to have a finite expansion by
\begin{equation}
\hat{T}=Th=\tilde{T}\prod_{i\in\{1,\dots,N_h\}\setminus\mathcal{S}}h_i.
\end{equation}

\subsubsection*{Expansion of the reformulated transformation law}
The transformation law \eq{DEQTrafoAlternativ} can now be rewritten entirely in terms of quantities with finite expansion
\begin{equation}
\label{DEQfiniteh}
-\hat{T}\textup{d}h+h\textup{d}\hat{T}-\hat{a}\hat{T}+\epsilon h\hat{T}\textup{d}\tilde{A}=0.
\end{equation}
Altogether, it was shown that for any solution $T$ of \eq{DEQTrafoAlternativ} there exists a solution $\hat{T}$ of \eq{DEQfiniteh} that has a finite expansion
\begin{equation}
\hat{T}=\sum_{n=l_\textup{min}}^{n_\textup{max}}\epsilon^n\hat{T}^{(n)}.
\end{equation}
Conversely, each solution $\hat{T}$ of \eq{DEQfiniteh} corresponds to a solution $T$ of \eq{DEQTrafoAlternativ} via $T=\hat{T}/h$. Thus, it can be avoided to calculate infinitely many coefficients in the expansion of $T$ by computing $\hat{T}$ instead. This can be done by expanding \eq{DEQfiniteh} in $\epsilon$:

\begin{equation}
-\hat{T}\textup{d}h+h\textup{d}\hat{T}=\sum_{n=2l_{\textup{min}}}^{n_\textup{max}+l_\textup{max}}\epsilon^n\sum_{k=l_\textup{min}}^{\textup{min}(l_\textup{max},n-l_\textup{min})}\left(-\textup{d}h^{(k)}\hat{T}^{(n-k)}+h^{(k)}\textup{d}\hat{T}^{(n-k)}\right),
\end{equation}

\begin{eqnarray}
\epsilon h\hat{T}\textup{d}\tilde{A}&=&\sum_{n=2l_{\textup{min}}}^{n_\textup{max}+l_\textup{max}}\epsilon^{n+1}\sum_{k=l_\textup{min}}^{\textup{min}(l_\textup{max},n-l_\textup{min})}h^{(k)}\hat{T}^{(n-k)}\textup{d}\tilde{A}\\
&=&\sum_{n=2l_{\textup{min}}+1}^{n_\textup{max}+l_\textup{max}+1}\epsilon^{n}\sum_{k=l_\textup{min}}^{\textup{min}(l_\textup{max},n-l_\textup{min}-1)}h^{(k)}\hat{T}^{(n-k-1)}\textup{d}\tilde{A},
\end{eqnarray}

\begin{equation}
\hat{a}\hat{T}=\sum_{n=l_{\textup{min}}}^{n_\textup{max}+k_\textup{max}}\epsilon^n\sum_{k=0}^{\textup{min}(k_\textup{max},n-l_\textup{min})}\hat{a}^{(k)}\hat{T}^{(n-k)}.
\end{equation}
Note that the equation at some order $k$ only involves $\hat{T}^{(n)}$ with $n\leq k$. Therefore, the $\hat{T}^{(n)}$ can be computed successively, starting with the lowest order. Given some $a(\varset)$, the first step is to calculate $h$ and $\hat{a}$, which fixes the values of $l_\textup{min}$, $l_\textup{max}$ and $k_\textup{max}$. The value of $n_\textup{max}$ remains unknown until the solution for $\hat{T}$ is known. Therefore, it is tested at each order $k$ whether $k=n_\textup{max}$. In order to do so, it has to be checked if $\hat{T}^{(n)}=0$ for all $n>k$ solves the equations of the remaining $\textup{max}(k_\textup{max}, l_\textup{max}+1)$ subsequent orders. The algorithm stops as soon as this test is successful and returns $T=\hat{T}/h$.

\subsection{Recursion over subsectors}
\label{sec:OffDiagPart}
The algorithm presented in subsection \ref{sec:ExpTrafo} is applicable to differential equations $a(\varset)$ of arbitrary dimension. However, if $a(\varset)$ comprises more than one sector, the computational cost can be significantly reduced by making use of its block-triangular form. In particular, the block-triangular form allows to compute the transformation to an $\epsilon$-form by means of a recursion over the subsectors of $a(\varset)$. Starting from the lowest subsector, at each step of the recursion the next diagonal block is transformed into $\epsilon$-form with the algorithm presented in subsection \ref{sec:ExpTrafo}. The off-diagonal blocks are transformed into $\epsilon$-form in a subsequent part of the recursion step, which will be the topic of this subsection. Similar considerations have been made in \cite{Caron-Huot:2014lda, Gehrmann:2014bfa, Lee:2014ioa}.

\subsubsection*{The interplay of overall and subsector transformations}

In order to investigate the recursion step, it is assumed that the first $p$ subsectors have already been transformed into a block-triangular $\epsilon$-form by a transformation $t_p$. Using the algorithm from subsection \ref{sec:ExpTrafo}, a transformation $t_{p+1}$ can be computed that transforms the next diagonal block into $\epsilon$-form. Up to this point, the transformation
\begin{equation}
t=\left(\,
\begin{array}{|ccc|c|}
\cline{1-4}
 & & &\\
\quad & t_p & \quad & 0 \\
 & & &\\ \hline
 & 0 & & t_{p+1}\\ \hline
\end{array}
\,\right)
\end{equation}
has been applied to the original $a(\varset)$. The intermediate expression $a_{I}$
\begin{equation}
a_{I}=t^{-1}at-t^{-1}\textup{d}t
\end{equation}
is of the form
\begin{equation}
\label{aBlockinEpsForm}
a_{I}=\left(\,
\begin{array}{|ccc|c|}
\cline{1-4}
 & & &\\
\quad &   \epsilon\tilde{c} & \quad &\,\, 0 \,\,\\
 & & &\\ \hline
 & b & & \epsilon\tilde{e}\\ \hline
\end{array}
\,\right),
\end{equation}
where $\tilde{c}$ and $\tilde{e}$ are in dlog-form with $\tilde{c}$ being block-triangular. The goal of this subsection is to devise an algorithm to compute the remaining transformation $t_r$, such that
\begin{equation}
\label{IntTrafoEq}
a^\prime=t_r^{-1}a_{I}t_r-t_r^{-1}\textup{d}t_r
\end{equation}
attains a block-triangular $\epsilon$-form:
\begin{equation}
a^\prime=\left(\,
\begin{array}{|ccc|c|}
\cline{1-4}
 & & &\\
\quad &   \epsilon\tilde{c}^\prime &\quad & 0 \\
 & & &\\ \hline
 &\epsilon\tilde{b}^\prime & & \epsilon\tilde{e}^\prime\\ \hline
\end{array}
\,\right).
\end{equation}
In subsection \ref{sec:GenProps} it has already been established that there is some freedom in the choice of the transformations $t_p$ and $t_{p+1}$. However, the algorithm from subsection \ref{sec:ExpTrafo} just returns one particular choice. It is conceivable that these independently made choices do not fit together and therefore have to be modified. It is thus important to investigate the relation between $t_p$ and $t_{p+1}$ and a transformation
\begin{equation}
T=\left(\,
\begin{array}{|ccc|c|}
\cline{1-4}
 & & &\\
 &   T_p & & 0 \\
 & & &\\ \hline
 & T_{p+1,p} & & T_{p+1}\\ \hline
\end{array}
\,\right)
\end{equation}
that transforms the full differential equation $a(\varset)$ into $\epsilon$-form. The superdiagonal block has been set to zero in order to preserve the block-triangular form. As the goal is only to find \emph{some} transformation, a subsequent invertible transformation $C$ and the multiplication by a rational function $g(\epsilon)$ may be chosen freely
\begin{equation}
\label{eq:tremdef}
t\cdot t_r=TCg(\epsilon).
\end{equation}
The relation between the diagonal blocks of $T$ and $t$ is assumed to be the following
\begin{eqnarray}
t_p&=&T_pc_pg_p(\epsilon),\\
t_{p+1}&=&T_{p+1}c_{p+1}g_{p+1}(\epsilon),
\end{eqnarray}
with invertible constant transformations $c_p$ and $c_{p+1}$.  Here $g_{p}(\epsilon)$ and $g_{p+1}(\epsilon)$ are matrices rational in $\epsilon$, which only encompass the additional degrees of freedom that are not already accounted for by the constant matrices $c_p$ and $c_{p+1}$. Note that this assumption is stronger than the mere assumption of the existence of $T$, since it has only been shown that a subsequent constant transformation and rescaling with a rational function will preserve the $\epsilon$-form of any differential equation. However, there also exist cases in which there is more freedom than that. A simple example is given by a differential equation with two sectors, where neither is a subsector of the other. Then there is no off-diagonal block and the integrals of the two sectors can be rescaled with different $\epsilon$-dependent rational functions without altering the $\epsilon$-form.

Let the blocks of the constant transformation $C$ be denoted by
\begin{equation}
C=\left(\,
\begin{array}{|ccc|c|}
\cline{1-4}
 & & &\\
 &   C_p & & C_{p,p+1} \\
 & & &\\ \hline
 & C_{p+1,p} & &C_{p+1}\\ \hline
\end{array}
\,\right).
\end{equation}
The freedom in the choice of $C$ can be used as follows
\begin{equation}
C_p=\,c_p,\quad C_{p+1}=\,c_{p+1},\quad C_{p+1,p}=0,\quad C_{p,p+1}=0.
\end{equation}
Again, the superdiagonal block has been set to zero in order to preserve the block-triangular form of the differential equation. Together with \eq{eq:tremdef} it follows that the computation of $t_r$ may be split into two consecutive steps by means of the following factorization:
\begin{equation}
t_r=t_Dt_{KJ},
\end{equation}
with 
\begin{equation}
t_D=\left(\,
\begin{array}{|ccc|c|}
\cline{1-4}
 & & &\\
\quad &   \mathbb{I} & \quad &\,\, 0 \,\,\\
 & & &\\ \hline
 & D & &\mathbb{I}\\ \hline
\end{array}
\,\right),\quad 
t_{KJ}=\left(\,
\begin{array}{|ccc|c|}
\cline{1-4}
 & & &\\
 &  K(\epsilon)  & & 0 \\
 & & &\\ \hline
 & 0 & & J(\epsilon) \\ \hline
\end{array}
\,\right),
\end{equation}
and
\begin{equation}
K\defeq g_p^{-1}g,\quad J\defeq g^{-1}_{p+1}g,\quad D\defeq Jc_{p+1}^{-1}T^{-1}_{p+1}T_{p+1,p}c_pK^{-1}.
\end{equation}
At this point it becomes apparent that in general $K$ and $J$ cannot be completely fixed with the choice of $g(\epsilon)$. Instead, both $K$ and $J$ need to be determined by \eq{IntTrafoEq}. The choice of $g$ can be used to fix one of the components of $K$ or $J$. The quantities $D$, $K$ and $J$ are determined by the following equations, which are implied by \eq{IntTrafoEq}
\begin{equation}
K^{-1}\tilde{c}K=\tilde{c}^\prime,\quad J^{-1}\tilde{e}J=\tilde{e}^\prime,
\end{equation}
\begin{equation}
\textup{d}D-\epsilon(\tilde{e}D-D\tilde{c})=b-\epsilon J\tilde{b}^\prime K^{-1}.
\end{equation}
In the latter equation, the product of three unknown quantities occurs in the term $\epsilon J\tilde{b}^\prime K^{-1}$. A linear ansatz for these quantities would result in nonlinear equations in the coefficients of the ansatz. This can be prevented by defining $b^\prime\defeq \epsilon J\tilde{b}^\prime K^{-1}$ and first solving
\begin{equation}
\label{DFullDEQ}
\textup{d}D-\epsilon(\tilde{e}D-D\tilde{c})=b-b^\prime
\end{equation}
for $D$ and $b^\prime$. Note that $b^\prime$ has to be in dlog-form, since $J$ and $K$ are independent of the invariants and therefore do not alter the dlog-form of $\tilde{b}^\prime$. 

\subsubsection*{The determination of $t_{KJ}$}
In a second step, the equations
\begin{equation}
K^{-1}\tilde{c}K=\tilde{c}^\prime,\quad J^{-1}\tilde{e}J=\tilde{e}^\prime, \quad b^\prime=\epsilon J\tilde{b}^\prime K^{-1},
\end{equation}
are solved, which is equivalent to finding a $t_{KJ}$ that transforms 
\begin{equation}
a^D(\varset)\defeq\left(\,
\begin{array}{|ccc|c|}
\cline{1-4}
 & & &\\
 & \epsilon\tilde{c}(\invariants)  & & 0 \\
 & & &\\ \hline
 &  b^\prime(\varset)  & & \epsilon\tilde{e}(\invariants) \\ \hline
\end{array}
\,\right)
\end{equation}
into $a^\prime$, which is in $\epsilon$-form. This can be achieved by a procedure outlined in \cite{Lee:2014ioa}, which is reproduced here for convenience. Since $a^D$ is in dlog-form, it can be written as
\begin{equation}
a^D=\sum_{l=1}^Na^D_l(\epsilon)\textup{d}\log(L_l(\invariants)).
\end{equation}
Every transformation $V(\epsilon)$ that transforms $a^D$ into $\epsilon$-form has to satisfy
\begin{equation}
\label{tepstoconst}
V(\epsilon)^{-1}\frac{a^D_l(\epsilon)}{\epsilon}V(\epsilon)=\tilde{o}_l,\quad l=1,\dots,N
\end{equation}
for constant matrices $\tilde{o}_l$. A necessary condition for $V(\epsilon)$ to exist is that the eigenvalues of $a^D_l(\epsilon)/\epsilon$ are constant. The following argument shows that this is indeed the case. Each of the $a^D_l$ is again of the same block-triangular form as $a^D$. The determinant of a block-triangular matrix equals the product of the determinants of its diagonal blocks. This leads to a factorization of the characteristic polynomials of the $a^D_l$
\begin{equation}
\det(a^D_l-\lambda\mathbb{I})=\det(\epsilon\tilde{c}_l-\lambda\mathbb{I})\det(\epsilon\tilde{e}_l-\lambda\mathbb{I}).
\end{equation}
In this form it is obvious that the eigenvalues of $a^D_l(\epsilon)$ are proportional to $\epsilon$. Therefore, the eigenvalues of $a^D_l(\epsilon)/\epsilon$ must be constant. In order to calculate such a transformation, \eq{tepstoconst} has to be solved. Since the constant matrices on the right-hand side are unknown, the components of $V(\epsilon)$ cannot be solved for directly. However, as the right-hand side of \eq{tepstoconst} is manifestly independent of $\epsilon$, the following holds
\begin{eqnarray}
V(\epsilon)^{-1}\frac{a^D_l(\epsilon)}{\epsilon}V(\epsilon)&=&V(\mu)^{-1}\frac{a^D_l(\mu)}{\mu}V(\mu)\\
\Leftrightarrow\frac{a^D_l(\epsilon)}{\epsilon}V(\epsilon)V(\mu)^{-1}&=&V(\epsilon)V(\mu)^{-1}\frac{a^D_l(\mu)}{\mu}\\
\Leftrightarrow\frac{a^D_l(\epsilon)}{\epsilon}V(\epsilon,\mu)&=&V(\epsilon,\mu)\frac{a^D_l(\mu)}{\mu},
\end{eqnarray}
with $V(\epsilon,\mu)\defeq V(\epsilon)V(\mu)^{-1}$. In the last form, for each $l=1,\dots,N$ there is a linear equation for $V(\epsilon,\mu)$. This set of equations can now be solved for the components of $V(\epsilon,\mu)$ subject to the constraint that the block-triangular form is preserved. Finally, a constant $\mu_0$ needs to be chosen such that $t_{KJ}= V(\epsilon,\mu_0)$ is non-singular. It is straightforward to check that this $t_{KJ}$ transforms $a^D$ into $\epsilon$-form:
\begin{eqnarray}
t_{KJ}^{-1}a^D_l(\epsilon)t_{KJ}
&=&V(\epsilon,\mu_0)^{-1} a^D_l(\epsilon)V(\epsilon,\mu_0)\\
&=&\epsilon V(\mu_0)V(\epsilon)^{-1}\frac{a^D_l(\epsilon)}{\epsilon}V(\epsilon)V(\mu_0)^{-1}\\
&=&\epsilon V(\mu_0)\tilde o_lV(\mu_0)^{-1}\\
&=&\epsilon\tilde{A}_l^\prime.
\end{eqnarray}

\subsubsection*{Setting up a recursion over subsectors for $t_{D}$}

In the following part of this section the determination of $t_D$ is considered. The goal is to find a rational $D$ and a $b^\prime$ in dlog-form that satisfy \eq{DFullDEQ}:
\begin{equation}
\label{DDEQ}
\textup{d}D-\epsilon(\tilde{e}D-D\tilde{c})=b-b^\prime.
\end{equation}
The block-triangular form of $\tilde{c}$ can be used to solve \eq{DDEQ} in a recursion over subsectors. To this end, all quantities are split according to the block-triangular structure into subsectors:
\begin{equation}
D=\left(D_1,\dots,D_p\right),\quad b=\left(b_1,\dots, b_p\right),\quad b^{\prime}=\left(b^{\prime}_1,\dots,b^{\prime}_p\right),
\end{equation}

\begin{align}
\tilde{c}&=\left(\,
\begin{array}{cccc}
\tilde{c}_{1} & & &   \\
\vdots & \ddots & & \\
 & & \tilde{c}_{p-1} & \\
\tilde{c}_{p,1} & \cdots & \tilde{c}_{p,p-1} & \tilde{c}_p\\ 
\end{array}
\,\right).
\end{align}
In this notation, \eq{DDEQ} may equivalently be written as a system of $p$ equations of the form
\begin{equation}
\textup{d}D_{k}-\epsilon(\tilde{e}D_{k}-D_{k}\tilde{c}_{k})=\left(b_{k}-\epsilon \sum_{i=k+1}^pD_i\tilde{c}_{i,k}\right)-b_{k}^\prime,\quad k=1,\dots, p.
\end{equation}
Note that the equation for a subsector $k$ only depends on the $D_n$ of higher subsectors $n \geq k$. It is therefore possible to solve for the $D_k$ in a recursion that starts with the highest subsector. As for the recursion step, suppose that the equations for the topmost $p-k$ subsectors have already been solved. The contribution of the higher subsectors to the equation of subsector $k$ is most naturally absorbed into the definition of
\begin{equation}
\bar{b}_k\defeq b_{k}-\epsilon  \sum_{i=k+1}^pD_i\tilde{c}_{i,k}.
\end{equation}
Thus, the following equation has to be solved 
\begin{equation}
\label{SubsecDDEQ}
\textup{d}D_k-\epsilon(\tilde{e}D_k-D_k\tilde{c}_k)=\bar{b}_k-b_k^\prime,
\end{equation}
with $\bar{b}_k$ being determined by the solution of the higher subsectors. 

\subsubsection*{Determination of the lowest order in the expansion of D}
The subsector index in \eq{SubsecDDEQ} is irrelevant for the following considerations and will be suppressed from now on. Since rational functions only posses poles of finite order, there exist finite integers $n_{\textup{min}}$ and $m_{\textup{min}}$ such that
\begin{equation}
\label{lowerBoundExists}
D=\sum_{m=m_{\textup{min}}}^\infty\epsilon^mD^{(m)},\quad \bar{b}=\sum_{n=n_{\textup{min}}}^\infty\epsilon^n\bar{b}^{(n)}.
\end{equation}
Consider the case $m_\textup{min}\leq n_\textup{min}$ and expand \eq{SubsecDDEQ} in $\epsilon$
\begin{align}
\textup{d}D^{(n_\textup{min})}-(\tilde{e}D^{(n_\textup{min}-1)}-D^{(n_\textup{min}-1)}\tilde{c}) &= \bar{b}^{(n_\textup{min})}-b^{\prime(n_\textup{min})}\\
\textup{d}D^{(n_\textup{min}-1)}-(\tilde{e}D^{(n_\textup{min}-2)}-D^{(n_\textup{min}-2)}\tilde{c}) &=-b^{\prime(n_\textup{min}-1)}\\
& \vdotswithin{=} \\
\textup{d}D^{(m_\textup{min}+1)}-(\tilde{e}D^{(m_\textup{min})}-D^{(m_\textup{min})}\tilde{c}) &=-b^{\prime(m_\textup{min}+1)}\\
\textup{d}D^{(m_\textup{min})} &=-b^{\prime(m_\textup{min})}.
\end{align}
Integrating the last equation yields
\begin{equation}
D^{(m_\textup{min})}=-\sum_{l=1}^N B_l^{\prime(m_\textup{min})}\log(L_l)+\textup{const},
\end{equation}
for constant matrices $B_l^{\prime(m_\textup{min})}$. Since $D$ is assumed to be rational, these matrices have to vanish and therefore it follows $b^{\prime(m_\textup{min})}=0$. This in turn implies $D^{(m_\textup{min})}=\textup{const}$. Proceeding to the next equation, it is evident that the term $(\tilde{e}D^{(m_\textup{min})}-D^{(m_\textup{min})}\tilde{c})$ is in dlog-form, since $D^{(m_\textup{min})}$ is constant and $\tilde{e}$ and $\tilde{c}$ are in dlog-form. By the same logic as before, this implies
\begin{align}
b^{\prime(m_\textup{min}+1)}&=\tilde{e}D^{(m_\textup{min})}-D^{(m_\textup{min})}\tilde{c},\\
D^{(m_\textup{min}+1)}&=\textup{const}.
\end{align}
The argument can only be repeated until the equation of order $n_\textup{min}-1$, since at higher orders also contributions from $\bar{b}$ appear. As the constant values of the $D^{(k)}$ with $k<n_\textup{min}-1$ do not affect the equations of the orders $n_\textup{min}$ or higher, they can be set to zero without loss of generality
\begin{align}
\label{DlowerBound}
D^{(k)}&=0,\quad \forall\, k<n_\textup{min}-1,\\
D^{(n_\textup{min}-1)}&=\textup{const},
\end{align}
which implies
\begin{equation}
\label{bprimelowerBound}
b^{\prime(k)}=0,\quad \forall\, k<n_\textup{min}.
\end{equation}
It has now been established that the $\epsilon$-expansion of $D$ can be assumed to start at order $n_\textup{min}-1$ or higher. Note that this assertion incorporates the case $m_\textup{min}>n_\textup{min}$ as well. Moreover, the coefficient at order $n_\textup{min}-1$ can be assumed to be constant.

\subsubsection*{Obtaining finite expansions}

The $\epsilon$-expansion of $D$ may still have infinitely many non-vanishing terms. Using ideas similar to those in subsection \ref{sec:ExpTrafo}, it will be shown that \eq{SubsecDDEQ} can be reformulated such that a solution for $D$ can be obtained by solving only finitely many differential equations. 

Since $D$ is assumed to be rational in $\epsilon$ and the invariants, a polynomial $f(\varset)$ has to exist such that $\check{D}\defeq Df$ has a finite $\epsilon$-expansion. Similarly, there exists a polynomial $k(\varset)$ such that $\check{b}\defeq \bar{b}k$ has a finite $\epsilon$-expansion as well. In order to fix $f$ and $k$ up to constant factors, both are required to only contain the minimal number of irreducible factors that are necessary to satisfy the aforementioned conditions. The products of all irreducible factors of $f$ and $k$ that are independent of the invariants are subsequently denoted by $\hat{f}$ and $\hat{k}$ respectively. Then their factorizations read
\begin{equation}
f(\varset)=\hat{f}(\epsilon)\prod_{i=1}^{N_f}\bar{f}_i(\varset),
\end{equation}
\begin{equation}
k(\varset)=\hat{k}(\epsilon)\prod_{i=1}^{N_k}\bar{k}_i(\varset).
\end{equation}
Furthermore, let $\gamma(\epsilon)$ be a polynomial with a minimal number of irreducible factors, such that $b^\prime\gamma$ has a finite expansion. Note that $\gamma(\epsilon)$ does not depend on the invariants since $b^\prime$ is in dlog-form.

For a given $\bar{b}$ it is straightforward to compute $k$, but as $D$ is not known in advance, $f$ cannot be calculated directly. Therefore, the relation between $f$ and $k$ has to be investigated. In order to do so, consider the \eq{SubsecDDEQ} rewritten in terms of $\check{D}$ 
\begin{equation}
\label{finite1DDEQ}
\sum_{i=1}^{N_f}\frac{-k\gamma\check{D}\textup{d}\bar{f}_i}{\bar{f}_i}=-k\gamma\textup{d}\check{D}+\epsilon k\gamma (\tilde{e}\check{D}-\check{D}\tilde{c})+f\gamma\check{b}-f\gamma k b^\prime.
\end{equation}
The right-hand side obviously has a finite expansion and thus also the left-hand side has to have a finite expansion. By similar arguments as in the previous subsection, each of the summands on the left-hand side has to have a finite expansion. Note that $\textup{d}\bar{f}_i/\bar{f}_i$ cannot be equal to a rational function with finite expansion. The same holds for $\check{D}/\bar{f}_i$ due to the minimality of $f$. Since $\gamma$ does not depend on the invariants, it follows that each $\bar{f}_i$ is equal to some $\bar{k}_j$ and thus
\begin{equation}
\label{eq:fInk}
k=\hat{k}(\epsilon)p(\varset)\bar{f}(\varset),
\end{equation}
with $p(\varset)$ being a polynomial and $\bar{f}$ denoting the product of all irreducible factors of $f$ that depend on the invariants. By applying this relation to \eq{finite1DDEQ} and dividing by $\bar{f}$, the following equation is obtained
\begin{equation}
\label{2dnformDcheckEq}
\sum_{i=1}^{N_f}\frac{-\hat{k}p\gamma\check{D}\textup{d}\bar{f}_i}{\bar{f}_i}=-\hat{k}p\gamma\textup{d}\check{D}+\hat{k}p\gamma\epsilon (\tilde{e}\check{D}-\check{D}\tilde{c})+\hat{f}\gamma\check{b}-\hat{f} k \gamma b^\prime.
\end{equation}
The same argument as above leads to $p(\varset)=r(\varset)\bar{f}(\varset)$ for some polynomial $r(\varset)$. Combining this relation with \eq{eq:fInk}, it is evident that the product $\bar{k}$ of all irreducible factors of $k$ that depend on the invariants contains two powers of $\bar{f}$
\begin{equation}
\label{kbarfsq}
\bar{k}=r\bar{f}^2.
\end{equation}
In order to learn about $\hat{f}$, the above equation is applied to \eq{2dnformDcheckEq} and subsequently divided by $\hat{f}$
\begin{equation}
\frac{\hat{k}r\gamma(\bar{f}\textup{d}\check{D}-\check{D}\textup{d}\bar{f}-\epsilon\bar{f}(\tilde{e}\check{D}-\check{D}\tilde{c}))}{\hat{f}}=\gamma\check{b}-\gamma k b^\prime.
\end{equation}
The irreducible factors of $r(\varset)$ cannot be equal to irreducible factors of $\hat{f}(\epsilon)$, because they are not independent of the invariants. The other factors in the numerator can be a product of an irreducible factor of $\hat{f}$ and a quantity with finite expansion. Since only $\hat{k}$ is known prior to solving the equations, some irreducible factors of $\hat{f}$ remain unknown.

\subsubsection*{Reformulation in terms of quantities with finite expansion}
Since $f$ cannot be used in practice, as it is not computable before solving for $D$, an alternative factor 
\begin{equation}
h(\varset)=\bar{h}(\varset)\hat{h}(\epsilon),
\end{equation}
will be defined such that the expansion of 
\begin{equation}
\hat{D}\defeq Dh,
\end{equation}
is finite. The minimality of $f$ implies that all irreducible factors of $f$ need to be irreducible factors of $h$ as well. The irreducible factors of $h$ that depend on both the invariants and $\epsilon$ can be defined by
\begin{equation}
\bar{h}(\varset)^2\defeq\bar{k}(\varset)s(\varset),
\end{equation}
where the polynomial $s(\varset)$ is required to have the minimal number of irreducible factors. By virtue of \eq{kbarfsq}, this definition ensures that $\bar{h}$ captures all irreducible factors of $\bar{f}$. 
As for the irreducible factors of $\hat{f}$, it is only known that some of them may be equal to irreducible factors of $\hat{k}$. The following definition incorporates all of these factors and leaves the missing factors to a factor $g(\epsilon)$ that has to be solved for 
\begin{equation}
\hat{h}(\epsilon)\defeq\hat{k}(\epsilon)g(\epsilon).
\end{equation}
Note that the minimality of $\hat{f}$ implies that $g(\epsilon)$ has a non-vanishing constant coefficient
\begin{equation}
g^{(0)}\neq 0.
\end{equation}

With the definitions $\hat{b}\defeq \bar{b}\bar{h}^2\hat{k}$ and $\hat{b}^\prime\defeq \hat{k}gb^\prime$, the differential equation \eq{SubsecDDEQ} can be rewritten entirely in terms of quantities with finite $\epsilon$-expansion
\begin{equation}
\label{FinalFormDDEQ}
-\textup{d}\bar{h}\hat{D}+\bar{h}\textup{d}\hat{D}-\epsilon\bar{h}(\tilde{e}\hat{D}-\hat{D}\tilde{c})=g(\epsilon)\hat{b}-\hat{b}^\prime\bar{h}^2.
\end{equation}
All quantities on the left-hand side have finite expansions by definition. The expansion of $\hat{b}$ must be finite as well, because $\hat{b}=\check{b}s$ and the expansion of $\check{b}$ is finite by definition. Together, this implies that $\hat{b}^\prime\bar{h}^2$ must have a finite expansion. Since $\hat{b}^\prime$ is in dlog-form, only factors that are independent of the invariants can render its expansion infinite. However, these factors could not be compensated by $\bar{h}^2$, which is a product of irreducible factors depending on both $\epsilon$ and the invariants, and therefore $\hat{b}^\prime$ itself has to have a finite expansion. Thus, all quantities in \eq{FinalFormDDEQ} indeed have a finite expansion. 

Altogether, the procedure is as follows: First, $\bar{k}$ and $\hat{k}$ are computed from the given $\bar{b}$, which then allows to infer $\bar{h}$ and $\hat{b}$. Subsequently, \eq{FinalFormDDEQ} can be solved for $\hat{D}$, $g$ and $\hat{b}^\prime$ all of which have a finite expansion. Finally, a solution of \eq{SubsecDDEQ} is obtained via $D=\hat{D}/(\bar{h}\hat{k}g)$.

\subsubsection*{Expansion of the reformulated equation for $t_{D}$}

As already mentioned above, the strategy to solve \eq{FinalFormDDEQ} is to expand it in $\epsilon$. The Taylor series of the polynomials $\bar{h}$, $\hat{k}$ and $g$ all start with a non-vanishing constant coefficient due to their minimality. This implies that the expansions of $\hat{D}$, $\hat{b}$ and $\hat{b}^\prime$ start at the same orders as those of $D$, $\bar{b}$ and $b^\prime$
\begin{equation}
\hat{D}=\sum_{m=n_\textup{min}-1}^{m_\textup{max}}\epsilon^m\hat{D}^{(m)},\quad\hat{b}=\sum_{p=n_\textup{min}}^{p_\textup{max}}\epsilon^p\hat{b}^{(p)},\quad \hat{b}^\prime=\sum_{s=n_\textup{min}}^{s_\textup{max}}\epsilon^s\hat{b}^{\prime(s)}.
\end{equation}
Let $\bar{h}_\textup{max}$, $g_\textup{max}\in\mathbb{Z}_{\geq 0}$ denote the highest non-vanishing order of the Taylor expansions of $\bar{h}$ and $g$ respectively. Expanding \eq{FinalFormDDEQ} in $\epsilon$ yields
\begin{eqnarray}
\sum_{n=n_\textup{min}-1}^{E_\textup{max}}\epsilon^nE^{(n)}=0,
\end{eqnarray}
with 
\begin{eqnarray}
E^{(n)}&=&\sum_{k=0}^{\textup{min}(\bar{h}_\textup{max},\,\,n-n_\textup{min}+1)}-\textup{d}\bar{h}^{(k)}\hat{D}^{(n-k)}+\bar{h}^{(k)}\textup{d}\hat{D}^{(n-k)}\\
&-&\sum_{k=0}^{\textup{min}(\bar{h}_\textup{max},\,\,n-n_\textup{min})}\bar{h}^{(k)}(\tilde{e}\hat{D}^{(n-k-1)}-\hat{D}^{(n-k-1)}\tilde{c})\\
&-&\sum_{k=n_\textup{min}}^{\textup{min}(p_\textup{max},\,\,n)}\hat{b}^{(k)}g^{(n-k)}\\
&+&\sum_{k=0}^{\textup{min}(2\bar{h}_\textup{max},\,\,n-n_\textup{min})}(\bar{h}^2)^{(k)}\hat{b}^{\prime(n-k)},
\end{eqnarray}
\begin{equation}
E_\textup{max}=\textup{max}(m_\textup{max}+\bar{h}_\textup{max}+1, p_\textup{max}+g_\textup{max}, 2\bar{h}_\textup{max}+s_\textup{max}).
\end{equation}
The equations $E^{(n)}=0$ are solved order by order, starting at the lowest order $n=n_\textup{min}-1$. Since $E_\textup{max}$ is unknown until the solution is known, it is tested at each order $n$ whether $n=E_\textup{max}$. To this end it is checked if
\begin{eqnarray}
\hat{D}^{(i)}&=&0,\quad i=n-\bar{h}_\textup{max},\dots,n,\\
g^{(i)}&=&0,\quad i=n-p_\textup{max}+1,\dots,n-n_\textup{min},\\
\hat{b}^{\prime(i)}&=&0,\quad i=n-2\bar{h}_\textup{max}+1,\dots,n.
\end{eqnarray}
Once this test has been successful, \eq{FinalFormDDEQ} is satisfied to all orders upon setting the coefficients of $\hat{D}$, $g$ and $\hat{b}^\prime$ of all, still undetermined, higher orders to zero too. The algorithm stops and returns $D=\hat{D}/(\bar{h}\hat{k}g)$.


\subsection{Leinartas decomposition}
\label{sec:Leinartas}
In the previous subsections it was shown that the computation of a transformation to $\epsilon$-form is equivalent to finding a rational solution of finitely many differential equations in the invariants. These equations do in general admit transcendental solutions as well. The strategy to find a rational solution of these equations is to make a rational ansatz. It is favorable to use an ansatz that depends linearly on its parameters, since this will translate linear differential equations into equations in the parameters that are linear again. 
This leaves the question which type of rational functions is sufficient to express any other rational function as a linear combination. An answer will be given in this subsection by showing that any rational function can be decomposed as a linear combination of a certain \emph{simple} type of rational functions.

In the univariate case, a partial fractions decomposition of the denominator polynomial could be used. However, in the multivariate case a naive generalization of partial fractioning may run into an infinite loop. This is illustrated by the following example
\begin{eqnarray}
\frac{1}{x(x+y)}&=&\frac{1}{xy}-\frac{1}{y(x+y)}\\
&=&\frac{1}{xy}-\left[\frac{1}{xy}-\frac{1}{x(x+y)}\right]=\frac{1}{x(x+y)}.
\end{eqnarray}
In the first equation, the partial fractions decomposition was applied with respect to $x$ and in the second equation it was applied with respect to $y$. Apparently, this procedure runs into a loop. This can be avoided by a more careful generalization of the partial fractioning procedure, as outlined in \cite{Lei78, 2012arXiv1206.4740R}. In the following, a brief account of this decomposition method is given, based on the above references and  \cite{Cox:2007:IVA:1204670}. The focus will be on the computational aspects and only those proofs will be shown that are relevant for the implementation of the decomposition. For the readers convenience, some definitions and standard results about polynomial rings that are used throughout this subsection are collected in appendix \ref{App:PolyAlgebra}.

\subsubsection*{Denominator decomposition}

Let $K[X]$ denote the ring of polynomials in $d$ variables $X=\{x_1,\dots,x_d\}$ with coefficients in a field $K$. Again, the cases $K=\mathbb{R}$ and $K=\mathbb{C}$ are the most relevant for the present application, but there is no need to specify the field for the following considerations.
\begin{defn}[Algebraic Independence]
A set of polynomials $\{f_1,\dots,f_m\}\subset K[X]$ is called algebraically independent if there exists no nonzero polynomial $\kappa$ in $m$ variables with coefficients in $K$ such that $\kappa(f_1,\dots,f_m)=0$ in $K[X]$. $\kappa$ is called annihilating polynomial.
\end{defn}
For the Leinartas decomposition, it is necessary to compute annihilating polynomials. Let $\{f_1,\dots,f_m\}\subset K[X]$ be a set of algebraically dependent polynomials and consider the ideal $I=\langle Y_1-f_1,\dots,Y_m-f_m\rangle\subseteq K[X,Y_1,\dots,Y_m]$. It is straightforward to check that the elements of the ideal $E=I\cap K[Y_1,\dots,Y_m]$  are annihilating polynomials. The following theorem provides a means to actually compute the elements of $E$.
\begin{thm}[Elimination Theorem]
Let $I\subset  K[X,Y_1,\dots,Y_m]$ be an ideal and $G$ be a Gr\"obner basis of $I$ with respect to lexicographic order with $X>Y_1>\dots >Y_m$. Then 
\begin{equation*}
G_Y=G\cap K[Y_1,\dots,Y_m]
\end{equation*}
is a Gr\"obner basis of the ideal $I\cap K[Y_1,\dots,Y_m]$.
\end{thm}
Thus, a Gr\"obner basis of $\langle Y_1-f_1,\dots,Y_m-f_m\rangle$ can be computed with standard algorithms \cite{GBBIB706, GBBIB699, Cox:2007:IVA:1204670} and the intersection of this basis with $K[Y_1,\dots,Y_m]$ gives a Gr\"obner basis for $E$. Every element of this basis is an annihilating polynomial.
\begin{lemma}
Any set of polynomials $\{f_1,\dots,f_m\}\subset K[X]$ with $m>d$ is algebraically dependent.
\end{lemma}
\begin{lemma}
\label{lem:anypower}
A finite set of polynomials $\{f_1,\dots,f_m\}\subset K[X]$ is algebraically dependent if and only if for all positive integers $e_1,\dots,e_m$ the set of polynomials $\{f_1^{e_1},\dots,f_m^{e_m}\}$ is algebraically dependent. 
\end{lemma}
The following considerations rely on a corollary of Hilbert's weak Nullstellensatz:
\begin{corollary}[Nullstellensatz certificate]
\label{lem:NSCertificate}
A finite set of polynomials $\{f_1,\dots, f_m\}\subset K[X]$ has no common zero in $\overline{K}^d$ if and only if there exist polynomials $h_1,\dots,h_m\in K[X]$ such that 
\begin{equation*}
1=\sum_{i=1}^mh_if_i.
\end{equation*}
The set of polynomials $\{h_1,\dots,h_m\}$ is called a Nullstellensatz certificate. 
\end{corollary}
A Nullstellensatz certificate is said to have degree $k$ if 
\begin{equation}
\max\{\textup{deg}(h_i)\,\,|\,\,i=1,\dots,m\}=k.
\end{equation}
Algorithm \ref{alg:Nullstellensatz} is a simple but sufficiently fast way to compute a Nullstellensatz certificate for a set of polynomials with no common zero.
\begin{algorithm}
\label{alg:Nullstellensatz}
 \KwIn{$\{f_1,\dots,f_m\}$ with no common zero.}
 \KwOut{Nullstellensatz certificate $\{h_1,\dots,h_m\}$ such that $\sum_{i=1}^mh_if_i=1$.}
 $k=0$\;
\Do{}{
  $\sum_{i=1}^mh_if_i=1$ with the $h_i$ being polynomials of degree $k$ with unknowns as coefficients. Extract a linear system of equations from this relation and solve it\;
  \eIf{solution exists}{
   \KwRet{certificate}
   }{
   $k=k+1;$
  }
 }
 \caption{Nullstellensatz certificate}
\end{algorithm}
The Leinartas decomposition is based on the following theorem, which provides a generalization of the partial fractions decomposition to the multivariate case.
\begin{thm}[Leinartas]
Let $f=p/q$ be a rational function with $p,q\in K[X]$ and $q=q_1^{e_1}\dots q_m^{e_m}$ be the unique factorization of $q$ in $K[X]$ and $V_i=\{x\in \overline{K}^d\, |\, q_i(x)=0\}$. Then $f$ can be written in the following form
\begin{equation*}
f=\sum_S\frac{p_S}{\prod_{i\in S}q_i^{b_i}}, \quad b_i\in \mathbb{N},\, p_S\in K[X],
\end{equation*}
with the sum running over all subsets $S\subseteq \{1, \dots, m\}$ with $\cap_{i\in S}V_i\neq\emptyset$ and $\{q_i \,|\, i\in S\}$ being algebraically independent.
\end{thm}
The proof of this theorem will be presented, because it directly translates to an algorithm that decomposes rational functions into the above form. The decomposition can be separated into two consecutive steps. In the first step, a form is attained that satisfies $\cap_{i\in S}V_i\neq\emptyset$ for each summand. This step is called \textit{Nullstellensatz decomposition}. Let $f=p/q$ be a rational function. In the case $\cap_{i=1}^mV_i\neq\emptyset$, the Nullstellensatz decomposition is already complete. Thus, it remains to consider the case $\cap_{i=1}^mV_i=\emptyset$. As $q_i$ has the same zero-set as $q_i^{e_i}$, it follows that $\{q_1^{e_1},\dots,q_m^{e_m}\}$ has no common zero in $\overline{K}^d$. According to corollary \ref{lem:NSCertificate}, a Nullstellensatz certificate $1=\sum_{i=1}^mh_iq^{e_i}_i$ exists in this situation. Multiplying the $f$ with this factor of one yields
\begin{equation}
f=\frac{p\sum_{i=1}^mh_iq^{e_i}_i}{q}=\sum_{i=1}^m\frac{ph_i}{q_1^{e_1}\cdots \widehat{q_i^{e_i}}\cdots q_m^{e_m}}.
\end{equation}
This step is applied repeatedly until the denominator factors of each term have a common zero. Note that this procedure will eventually stop since single irreducible factors always have a zero $V_i\neq \emptyset$. In the second step, the goal is to achieve that $\{q_1,\dots,q_m\}$ is algebraically independent for each summand. Let $f=p/q$ be a summand of the Nullstellensatz decomposition. If $\{q_1,\dots,q_m\}$ is algebraically independent, then this term is already in the desired form. If this is not the case, the set $\{q_1^{e_1},\dots,q_m^{e_m}\}$ is also algebraically dependent by virtue of lemma \ref{lem:anypower}. Therefore, an annihilating polynomial $\kappa=\sum_{\nu\in S}c_\nu Y^\nu\in K[Y_1,\dots,Y_m]$ exists, which has been written in multi-index notation with $S\subset\mathbb{N}^m$. Let $\mu\in S$ refer to the powers of the monomial with the smallest norm $\|\mu\|=\sum_{i=1}^m\mu_i$. The annihilating polynomial vanishes on $Q=(q_1^{e_1},\dots,q_m^{e_m})$ 
\begin{eqnarray}
\kappa(Q)&=&0\\
\Rightarrow \quad c_\mu Q^\mu & = & -\sum_{\nu\in S\setminus\{\mu\}}c_\nu Q^\nu\\
\Rightarrow \quad 1 & = & \frac{-\sum_{\nu\in S\setminus\{\mu\}}c_\nu Q^\nu}{c_\mu Q^\mu}.
\end{eqnarray}
This factor of one can be used to decompose $f$
\begin{equation}
f=\frac{p}{q}= \sum_{\nu\in S\setminus\{\mu\}}\frac{-pc_\nu Q^\nu}{c_\mu Q^{\mu+1}}=\sum_{\nu\in S\setminus\{\mu\}}\frac{-pc_\nu}{c_\mu}\prod_{i=1}^m\frac{q_i^{e_i\nu_i}}{q_i^{e_i(\mu_i+1)}}.
\end{equation}
As $\mu$ has the smallest norm in $S$, there has to exists some $j$ for each $\nu\in S$ such that $\mu_j+1\leq \nu_j$ and therefore $e_j(\mu_j+1)\leq e_j\nu_j$. So in each summand at least one factor in the denominator cancels. Again, this step is applied repeatedly to all summands whose denominator factors are algebraically dependent. Eventually, this procedure will stop, since a single irreducible factor is obviously algebraically independent. This completes the proof of the Leinartas theorem. Following this proof, a recursive algorithm can be built that computes the above decomposition of rational functions.

\subsubsection*{Numerator decomposition}

The Leinartas decomposition as presented in \cite{Lei78, 2012arXiv1206.4740R} leaves the numerator polynomial untouched. However, by employing multivariate polynomial division, the above decomposition can be extended to the numerator polynomial as well, which results in summands with simpler numerator polynomials. The precise meaning of simple in this context will be stated below. 

Consider a summand $f=p/(q_1^{e_1}\dots q_m^{e_m})$ of the above decomposition, i.e. with $\cap_{i\in S}V_i\neq\emptyset$ and the $q_i$ being algebraically independent. The numerator polynomial $p$ can be decomposed according to the following theorem (cf. \cite{Cox:2007:IVA:1204670}).
\begin{thm}[Division Algorithm]
\label{thm:DivAlg}
Fix some monomial ordering on $\mathbb{Z}_{\geq 0}^d$ and let $(f_1,\dots,f_m)$ be an ordered m-tuple of polynomials in $K[X]$. Then every $p\in K[X]$ can be written as
\begin{equation*}
p=\beta_1f_1+\dots+\beta_mf_m+r
\end{equation*}
with $\beta_1,\dots,\beta_m,r\in K[X]$ and either $r=0$ or $r$ is a linear combination of monomials with coefficients in $K$ such that no monomial is divisible by any of the $\textup{LT}(f_1),\dots,\textup{LT}(f_m)$. Moreover, for all $\beta_if_i\neq 0$ the following holds
\begin{equation*}
\textup{multideg}(p)\geq\textup{multideg}(\beta_if_i).
\end{equation*}
\end{thm}
It should be noted that the resulting decomposition depends on both the ordering of the $(f_1, \dots, f_m)$ and the monomial ordering. Let the ordered tuple of polynomials be given by the set of denominator polynomials $(q_1,\dots,q_m)$ and apply the above theorem to the numerator polynomial
\begin{equation}
p=\beta_1q_1+\dots+\beta_mq_m+r,
\end{equation}
to arrive at
\begin{equation}
\label{fNumDecomposition}
f=\frac{r}{q_1^{e_1}\dots q_m^{e_m}}+\sum_{i=1}^m\frac{\beta_i}{q_1^{e_1}\dots q_i^{e_i-1}\dots q_m^{e_m}}.
\end{equation}
The denominator factors of the resulting summands are still algebraically independent, since every subset of an algebraically independent set of polynomials is algebraically independent. Moreover, every subset of a set of polynomials that share a common zero, has a common zero as well. So after decomposing the numerator as above, the denominator polynomials of the resulting summands still have a common zero and are algebraically independent. Therefore, this decomposition can be applied recursively. The recursion stops at a summand whenever there is no monomial of the numerator polynomial that is divisible by the leading term of any of the denominator polynomials. It has to be shown that the recursion will always stop after a finite number of steps. For the first summand in \eq{fNumDecomposition} the recursion trivially stops. Concerning the other terms, it is sufficient to show that the multidegree strictly decreases
\begin{equation}
\label{eq:multidegStrictlydec}
\textup{multideg}(p)>\textup{multideg}(\beta_i)
\end{equation}
at each step, due to property \ref{def:monOrd:3} of definition \ref{def:monOrd} given in appendix \ref{App:PolyAlgebra}. Lemma \ref{lem:lowerboundmonoordering} implies $\textup{multideg}(q_i)\geq 0$ with respect to any monomial ordering. However, the case $\textup{multideg}(q_i)=0$ cannot occur, since it implies $q_i=const$. Thus, $\textup{multideg}(q_i)$ is strictly greater than zero. Using property \ref{def:monOrd:2} of definition \ref{def:monOrd} and lemma \ref{lem:multidegProperties} it follows
\begin{equation}
\textup{multideg}(\beta_iq_i)=\textup{multideg}(q_i)+\textup{multideg}(\beta_i)>\textup{multideg}(\beta_i).
\end{equation}
Theorem \ref{thm:DivAlg} implies $\textup{multideg(p)}\geq\textup{multideg}(\beta_iq_i)$, which together with the above inequality proves \eq{eq:multidegStrictlydec}. This completes the decomposition of the numerator polynomial. 

The terms in such a decomposition are not necessarily linearly independent over $K$, as the following example illustrates
\begin{equation}
\frac{1}{x+y}+\frac{y}{x(x+y)}-\frac{1}{x}=0.
\end{equation}
In the last step, this redundancy is removed by eliminating all such relations from the set of summands. Altogether, it has been demonstrated that every multivariate rational function can be decomposed into $K$-linearly independent summands such that denominator polynomials of each summand share a common zero and are algebraically independent, and the numerator polynomial is not divisible by the leading term of any of its denominator polynomials. In the following this decomposition is referred to as \emph{Leinartas decomposition} and the individual summands are said to be in \emph{Leinartas form}.

\subsection{Solving for a rational transformation}
\label{sec:Solving}
In this subsection the Leinartas decomposition of multivariate rational functions will be employed to solve the differential equations that appear at each order of the expansion of \eq{DEQfiniteh} and \eq{FinalFormDDEQ}. In both cases these differential equations, in general, admit transcendental solutions for $\hat{T}^{(n)}$ and $\hat{D}^{(n)}$. Since only rational solutions are of interest for the present application, it suggests itself to solve these equations with a  rational ansatz. In the previous subsection it has been shown that any multivariate rational function can be written as a linear combination of rational functions in Leinartas form. In particular, this implies that the rational solutions of the differential equations above can also be written as a linear combination of these functions. Therefore, the ansatz can be chosen to be a linear combination of rational functions in Leinartas form with unknown coefficients without losing generality.

\subsubsection*{The strategy for diagonal blocks}
First, consider the part of the algorithm for the diagonal-blocks, outlined in subsection \ref{sec:ExpTrafo}. The following ansatz is used for each Taylor coefficient of $\hat{T}$
\begin{equation}
\hat{T}^{(n)}=\sum_{k=1}^{|\mathcal{R}_T|}\tau_k^{(n)}r_k(\invariants),
\end{equation}
\begin{equation}
\mathcal{R}_T\defeq \left\{r_1(\invariants),\dots,r_{|\mathcal{R}_T|}(\invariants)\right\},
\end{equation}
where the $\tau_k^{(n)}$ denote $m \times m$ matrices of unknown parameters. It is necessary to determine the right set $\mathcal{R}_T$ of rational functions in Leinartas form that is sufficiently large to encompass a solution. To this end, it is very useful that the determinant of $\hat{T}$ can easily be computed by virtue of \eq{DetIsFixed}. The powers of the irreducible factors in the determinant can be used as input for a heuristic procedure to generate an ansatz, which will be described in detail in a future publication \cite{Meyer:0000abc}.

Note that $\textup{d}\tilde{A}$ is also unknown in \eq{DEQfiniteh}. However, the dependence of $\tilde{A}$ on the invariants is restricted by the requirement that $\textup{d}\tilde{A}$ is in dlog-form
\begin{equation}
\tilde{A}=\sum_{l=1}^N\alpha_l \log(L_l(\invariants)),
\end{equation}
where the $\alpha_l$ are considered to be $m \times m$ matrices of unknown parameters. The set of letters  
\begin{equation}
\mathcal{A}=\{L_1(\invariants),\dots,L_N(\invariants)\}
\end{equation}
has to be chosen such that it contains all letters that are necessary for a resulting $\epsilon$-form. A natural choice is to take the set of all irreducible denominator factors occurring in $\hat{a}$. In subsection \ref{sec:GenProps} it was shown that \eq{TraceIsFixed} fixes the traces of all $\alpha_l$ and thereby reduces the number of free parameters that have to be solved for. 

Upon inserting this ansatz in the expansion of \eq{DEQfiniteh} and requiring the resulting equations to hold for all allowed values of the invariants, a system of equations in the unknown parameters is obtained. It is possible that $\hat{T}^{(n)}$ is not fully determined by the equations of order $n$ or lower. If $\tilde{A}$ is not fully determined by these equations as well, it may happen that terms, which are nonlinear in the parameters, arise in the equations of order $n+1$. This is due to the term $\epsilon T\textup{d}\tilde{A}$ in \eq{DEQTrafoAlternativ}. Therefore, the system of equations in the unknown parameters is, in general, polynomial.

\subsubsection*{The strategy for off-diagonal blocks}

A similar strategy is employed for the part of the algorithm that is concerned with the off-diagonal blocks, which is discussed in subsection \ref{sec:OffDiagPart}. For the coefficients of $\hat{D}$ in the expansion of \eq{FinalFormDDEQ} the ansatz
\begin{equation}
\hat{D}^{(n)}=\sum_{k=1}^{|\mathcal{R}_D|}\delta_k^{(n)}r_k(\invariants),
\end{equation}
\begin{equation}
\mathcal{R}_D\defeq \left\{r_1(\invariants),\dots,r_{|\mathcal{R}_D|}(\invariants)\right\},
\end{equation}
is used, where the $\delta_k^{(n)}$ are matrices of unknown parameters of the same dimensions as $\hat{D}$ and $\mathcal{R}_D$ denotes a set of rational functions in Leinartas form. The coefficients of $\hat{b}^\prime$ are unknown, but assumed to be in dlog-form 
\begin{equation}
\hat{b}^{\prime(n)}=\sum_{l=1}^N\beta^{(n)}_l\textup{d}\log\left(L_l(\invariants)\right),
\end{equation}
where the $\beta^{(n)}_l$ denote matrices of unknown parameters. The set of letters is taken to be the set of irreducible denominator factors in $\hat{b}$. Since the constant coefficient $g^{(0)}$ of $g(\epsilon)$ is nonzero, \eq{FinalFormDDEQ} can be divided by $g^{(0)}$. Subsequently, this factor can be absorbed into the definitions of $\hat{D}$ and $\hat{b}^\prime$. Effectively, this amounts to setting $g^{(0)}=1$ without loss of generality. All higher Taylor coefficients of $g(\epsilon)$ are treated as unknown parameters. Once all of the above is inserted into the expansion of \eq{FinalFormDDEQ}, linear equations in the unknown parameters are obtained at each order.

\subsubsection*{Beyond the canonical form}
The presented algorithm is able to compute a rational transformation of a given differential equation into $\epsilon$-form, whenever such a transformation exists and it is decomposable in terms of the ansatz that is used. If no such transformation exists for the given ansatz, the equations in the parameters of the ansatz will not have a solution. In this case, either the ansatz is not general enough or a rational transformation to $\epsilon$-form does not exist at all. A sufficient condition for the latter case is the presence of non-rational factors in the determinant of the transformation $T$, which can be computed with \eq{DetIsFixed}. In this case an $\epsilon$-form may still be attainable with a non-rational transformation. 

However, it is well known \cite{Caffo:1998du, Laporta:2004rb, Bloch:2013tra, Adams:2014vja, Bloch:2014qca, Adams:2015gva, Bloch:2016izu, Remiddi:2016gno, Adams:2016xah, Bonciani:2016qxi} that Feynman integrals exist that satisfy higher order differential equations and therefore a canonical form as in \eqref{EpsForm} can not exist for these integrals. It has been observed that for the differential equations of these integrals a dlog-form with linear dependence on $\epsilon$ can be attained
\begin{equation}
\label{linEpsForm}
A(\varset)=\sum_{l=1}^N(\bar{A}_l+\epsilon\tilde{A}_l)\log(L_l(\invariants)),
\end{equation}
where the $\bar{A}_l$ and $\tilde{A}_l$ denote constant matrices. In this more general case, the transformation law \eqref{DEQTrafoAlternativ} generalizes as follows
\begin{equation}
\label{genDEQTrafo}
\textup{d}T-aT+\epsilon T\textup{d}\tilde{A}=-T\textup{d}\bar{A}.
\end{equation}
Note that the term on the right-hand side has not been present in the original transformation law \eqref{DEQTrafoAlternativ}. The main ideas of the presented algorithm carry over to the problem of finding a transformation that satisfies the more general equation \eq{genDEQTrafo}. Since \eq{genDEQTrafo} is invariant under the multiplication of $T$ with a rational function $g(\epsilon)$, a procedure similar to the one described in section \ref{sec:Algorithm} can be used to construct a transformation with finite expansion. By expanding \eqref{genDEQTrafo} in $\epsilon$ and making the ansatz for $\textup{d}\bar{A}$ in the same way as for $\textup{d}\tilde{A}$, the algorithm generalizes naturally to this more general situation. If no canonical form exists, there will be no solution with $\bar{A}=0$. In this case there may still be a solution with non-vanishing $\bar{A}$, which then corresponds to the more general form \eq{linEpsForm} of the differential equation.

\section{Applications}
\label{sec:Applications}
In this section, the algorithm described in the previous section is applied to a set of non-trivial examples. These are given by four two-loop double box topologies, which can be specified by seven propagators and two irreducible scalar products:
\begin{equation}
I(\nu_1,\dots,\nu_9)=\int\frac{\textup{d}^dl_1}{\ima\pi^{d/2}}\frac{\textup{d}^dl_2}{\ima\pi^{d/2}}\frac{P_8^{-\nu_8}P_9^{-\nu_9}}{P_1^{\nu_1}\dots P_7^{\nu_7}}.
\end{equation}
A finite basis of master integrals has been computed for each of the examples with Reduze \cite{Studerus:2009ye, vonManteuffel:2012np}.

\subsection{Two loop single top-quark production}
The integrals considered in this example are necessary to include certain color suppressed contributions in the NNLO QCD corrections to single top-quark production \cite{Assadsolimani:2014oga}, which have been neglected a previous calculation \cite{Brucherseifer:2014ama}. These integrals have not been considered before and therefore represent a new result. 

The algorithm is applied to the planar topology 1 and the non-planar topology 2, which are given by the following sets of propagators.
\paragraph*{Topology 1:}
\begin{equation}
\begin{array}{lll}
P_1=l_2^2, & P_4=(l_2+p_2)^2, & P_7=(l_1+l_2-p_1+p_3)^2, \\
P_2=l_1^2-m_W^2, & P_5=(l_1-p_4)^2, &  P_8=(l_1-p_2)^2, \\
P_3=(l_1+p_3)^2, &P_6=(l_2-p_1)^2, & P_9=(l_2+p_3+p_1)^2. \\  
\end{array}
\end{equation}
\paragraph*{Topology 2:}
\begin{equation}
\begin{array}{lll}
P_1=l_2^2, & P_4=(l_2-p_2)^2, & P_7=(l_1-l_2-p_1+p_3)^2, \\
P_2=l_1^2-m_W^2, & P_5=(l_1-p_4)^2, &  P_8=(l_1+p_2)^2, \\
P_3=(l_1+p_3)^2, &P_6=(l_2-l_1-p_3)^2, & P_9=(l_2-p_3)^2. \\  
\end{array}
\end{equation}
The momenta $p_1$ and $p_2$ are counted incoming and $p_3$ and $p_4$ are counted outgoing. Both topologies are expressed using the following invariants
\begin{equation}
p_1^2=0,\quad p_2^2=0,\quad p_3^2=0,\quad p_4^2=m_t^2,
\end{equation}
\begin{equation}
p_1+p_2=p_3+p_4,
\end{equation}
\begin{equation}
s\defeq(p_1+p_2)^2,\quad t\defeq(p_2-p_3)^2.
\end{equation}
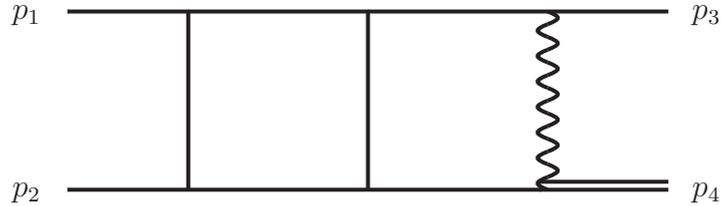
\begin{figure}[ht]
\begin{center}
\fcolorbox{white}{white}{
  \begin{picture}(330,120) (-5,-70)
   \SetScale{0.5}
   \unitlength=0.5pt
    \SetWidth{3.0}
    \SetColor{Black}
    \Line(96,51)(544,51)
    \Line[double,sep=6](448,-80)(544,-80)
    \Photon(454,-83)(454,51){7.5}{7}
    \Line(96,-83)(449,-83)
    \Text(564,-93)[lb]{\large{\Black{$p_4$}}}
    \Text(564,41)[lb]{\large{\Black{$p_3$}}}
    \Text(55,41)[lb]{\large{\Black{$p_1$}}}
    \Text(55,-93)[lb]{\large{\Black{$p_2$}}}
    \Line(186,51)(186,-83)
    \Line(320,51)(320,-83)
  \end{picture}
}
\end{center}
\caption{Two loop graph of the planar topology 1.}
\end{figure}
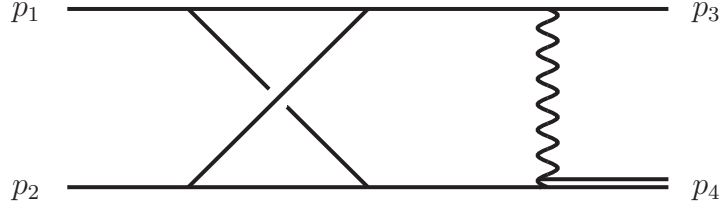
\begin{figure}[ht]
\begin{center}
\fcolorbox{white}{white}{
  \begin{picture}(330,120) (-5,-70)
   \SetScale{0.5}
   \unitlength=0.5pt
    \SetWidth{3.0}
    \SetColor{Black}
    \Line(96,51)(544,51)
    \Line[double,sep=6](448,-80)(544,-80)
    \Photon(454,-83)(454,51){7.5}{7}
    \Line(96,-83)(449,-83)
    \Text(564,-93)[lb]{\large{\Black{$p_4$}}}
    \Text(564,41)[lb]{\large{\Black{$p_3$}}}
    \Text(55,41)[lb]{\large{\Black{$p_1$}}}
    \Text(55,-93)[lb]{\large{\Black{$p_2$}}}
    \Line(186,51)(246.3,-9.3)
    \Line(259.7,-22.7)(320,-83)
    \Line(320,51)(186,-83)
  \end{picture}
}
\end{center}
\caption{Two loop graph of the non-planar topology 2}
\end{figure}
The integration-by-parts reduction of topology 1 to master integrals reveals that it admits a basis of 31 master integrals: 
\begin{equation}
\label{OldBasisSTT1}
\begin{array}{lll}
\vec{g}^{\,\textup{t1}}(\epsilon,s,t,m_t^2,m_W^2)\,\,\,=&\big(I_\textup{t1}(0,1,0,1,0,1,0,0,0),&I_\textup{t1}(0,1,0,1,1,1,0,0,0),\\
&I_\textup{t1}(0,0,1,1,1,1,0,0,0),&I_\textup{t1}(0,1,1,1,1,1,0,0,0),\\
&I_\textup{t1}(1,1,0,0,0,0,1,0,0),&I_\textup{t1}(1,1,-1,0,0,0,1,0,0),\\
&I_\textup{t1}(0,1,0,1,0,0,1,0,0),&I_\textup{t1}(-1,1,0,1,0,0,1,0,0),\\
&I_\textup{t1}(0,0,1,1,0,0,1,0,0),&I_\textup{t1}(0,1,1,1,0,0,1,0,0),\\
&I_\textup{t1}(1,1,1,1,0,0,1,0,0),&I_\textup{t1}(1,1,1,1,-1,0,1,0,0),\\
&I_\textup{t1}(1,1,0,0,1,0,1,0,0),&I_\textup{t1}(1,0,1,0,1,0,1,0,0),\\
&I_\textup{t1}(1,1,1,0,1,0,1,0,0),&I_\textup{t1}(1,1,1,-1,1,0,1,0,0),\\
&I_\textup{t1}(0,1,0,0,0,1,1,0,0),&I_\textup{t1}(0,1,0,1,0,1,1,0,0),\\
&I_\textup{t1}(-1,1,0,1,0,1,1,0,0),&I_\textup{t1}(1,1,0,1,0,1,1,0,0),\\
&I_\textup{t1}(1,1,-1,1,0,1,1,0,0),&I_\textup{t1}(0,1,1,1,0,1,1,0,0),\\
&I_\textup{t1}(0,1,0,0,1,1,1,0,0),&I_\textup{t1}(-1,1,0,0,1,1,1,0,0),\\
&I_\textup{t1}(1,1,0,0,1,1,1,0,0),&I_\textup{t1}(1,1,-1,0,1,1,1,0,0),\\
&I_\textup{t1}(0,1,0,1,1,1,1,0,0),&I_\textup{t1}(1,1,0,1,1,1,1,0,0),\\
&I_\textup{t1}(1,1,1,1,1,1,1,0,0),&I_\textup{t1}(1,1,1,1,1,1,1,0,-1),\\
&I_\textup{t1}(1,1,1,1,1,1,1,-1,0)\big).&
\end{array}
\end{equation}
Similarly, the scalar integrals of topology 2 can be expressed as linear combinations of the following 35 master integrals:
\begin{equation}
\label{OldBasisSTT2}
\begin{array}{lll}
\vec{g}^{\,\textup{t2}}(\epsilon,s,t,m_t^2,m_W^2)\,\,\,=&\big(I_\textup{t2}(1,1,0,0,0,1,0,0,0),&I_\textup{t2x12}(-1,1,0,1,0,1,0,0,0),\\
&I_\textup{t2x12}(0,1,0,1,0,1,0,0,0),&I_\textup{t2}(-1,1,0,1,0,1,0,0,0),\\
&I_\textup{t2}(0,1,0,1,0,1,0,0,0),&I_\textup{t2}(1,0,0,0,1,1,0,0,0),\\
&I_\textup{t2}(1,1,-1,0,1,1,0,0,0),&I_\textup{t2}(1,1,0,0,1,1,0,0,0),\\
&I_\textup{t2x12}(0,1,0,1,1,1,0,0,0),&I_\textup{t2}(0,1,0,1,1,1,0,0,0),\\
&I_\textup{t2x12}(1,1,-1,1,1,1,0,0,0),&I_\textup{t2x12}(1,1,0,1,1,1,0,0,0),\\
&I_\textup{t2}(1,1,-1,1,1,1,0,0,0),&I_\textup{t2}(1,1,0,1,1,1,0,0,0),\\
&I_\textup{t2}(0,0,1,1,1,1,0,0,0),&I_\textup{t2x12}(-1,1,1,1,1,1,0,0,0),\\
&I_\textup{t2x12}(0,1,1,1,1,1,0,0,0),&I_\textup{t2}(-1,1,1,1,1,1,0,0,0),\\
&I_\textup{t2}(0,1,1,1,1,1,0,0,0),&I_\textup{t2}(-1,1,0,1,0,0,1,0,0),\\
&I_\textup{t2}(0,1,-1,1,0,0,1,0,0),&I_\textup{t2}(0,1,1,1,0,0,1,0,0),\\
&I_\textup{t2x12}(1,1,1,1,-1,0,1,0,0),&I_\textup{t2x12}(1,1,1,1,0,0,1,0,0),\\
&I_\textup{t2}(1,1,1,1,-1,0,1,0,0),&I_\textup{t2}(1,1,1,1,0,0,1,0,0),\\
&I_\textup{t2}(1,1,-1,1,0,1,1,0,0),&I_\textup{t2}(1,1,0,1,0,1,1,0,0),\\
&I_\textup{t2}(1,1,1,1,0,1,1,0,0),&I_\textup{t2}(1,1,-1,1,1,1,1,0,0),\\
&I_\textup{t2}(1,1,0,1,1,1,1,0,0),&I_\textup{t2}(1,0,1,1,1,1,1,0,0),\\
&I_\textup{t2}(1,1,1,1,1,1,1,-2,0),&I_\textup{t2}(1,1,1,1,1,1,1,-1,0),\\
&I_\textup{t2}(1,1,1,1,1,1,1,0,0)\big).&
\end{array}
\end{equation}
Here the subscript $\textup{t2x12}$ refers to the set of propagators of the topology 2 with the momenta $p_1$ and $p_2$ exchanged. In order to work with dimensionless integrals, the mass-dimension of the master integrals is factored out by defining
\begin{equation}
f_i(\epsilon,x,y,z)=(m_W)^{-\textup{dim}(g_i)}g_i(\epsilon,s,t,m_t,m_W).
\end{equation}
The dimensionless integrals only depend on the dimensionless parameters 
\begin{equation}
x=\frac{s}{m_W^2},\quad y=\frac{t}{m_W^2},\quad z=\frac{m_t^2}{m_W^2}.
\end{equation}
The canonical bases of both topologies have been computed by means of an implementation of the algorithm from section  \ref{sec:Algorithm} in \texttt{Mathematica}. The transformations are presented by showing the decomposition of the integrals of the original bases with respect to the canonical bases. For brevity, only three integrals of each topology are shown in the following, the full results are provided in ancillary files accompanying the arXiv preprint.

\paragraph*{Topology 1:}

\footnotesize
\begin{align}
f^\textup{t1}_{10}=&
\left(-\frac{\epsilon^2 \left((9\epsilon-3) x^2+x (-\epsilon (23 z+5)+7 z+1)+2 z (7\epsilon z+\epsilon-2 z)\right)}{2 (2\epsilon-1)^2 (3\epsilon-1) (x-z)^2}\right)f^{\textup{t1}\prime}_{7}\nonumber\\
&+\left(\frac{\epsilon^2 \left((3\epsilon-1) x^2+x (-5\epsilon (z-1)+z-1)+2\epsilon (z-1) z\right)}{2 (2\epsilon-1)^2 (3\epsilon-1) (x-z)^2}\right)f^{\textup{t1}\prime}_{8}\nonumber\\
&+\left(\frac{\epsilon^2 x (\epsilon (5 x-5 z+3)-x+z-1)}{2 (2\epsilon-1)^2 (3\epsilon-1) (x-z)^2}\right)f^{\textup{t1}\prime}_{9}
+\left(-\frac{\epsilon^2 x (x-z+1)}{2 (2\epsilon-1)^2 (x-z)^2}\right)f^{\textup{t1}\prime}_{10},\nonumber\\
\end{align}
\normalsize

\footnotesize
\begin{align}
f^\textup{t1}_{18}=&
\left(\frac{\epsilon (2\epsilon (x-z-2)+1)}{(2\epsilon-1) (3\epsilon-1) (x-z)}\right)f^{\textup{t1}\prime}_{1}
+\left(\frac{\epsilon (\epsilon (x-z-7)+2)}{(2\epsilon-1) (3\epsilon-1) (x-z)}\right)f^{\textup{t1}\prime}_{7}\nonumber\\
&+\left(-\frac{\epsilon^2 (x-z+1)}{(2\epsilon-1) (3\epsilon-1) (x-z)}\right)f^{\textup{t1}\prime}_{8}
+\left(-\frac{3\epsilon}{(2\epsilon-1) (x-z)}\right)f^{\textup{t1}\prime}_{17}\nonumber\\
&+\left(\frac{\epsilon}{(2\epsilon-1) (x-z)}\right)f^{\textup{t1}\prime}_{18}
+\left(\frac{\epsilon^2 (x-z+1)}{(2\epsilon-1) (3\epsilon-1) (x-z)}\right)f^{\textup{t1}\prime}_{19},
\end{align}
\normalsize

\footnotesize
\begin{align}
f^\textup{t1}_{31}=&
\left(\frac{x-y-z}{2 x (x-z) (x+y-z)}\right)f^{\textup{t1}\prime}_{1}
+\left(\frac{2 x-z}{2 x^2 (x-z)}\right)f^{\textup{t1}\prime}_{3}
+\left(\frac{2 x^2+x (3 y-2 z)+(y-1) (y-z)}{2 x^2 (x+y-z) (x+y-z+1)}\right)f^{\textup{t1}\prime}_{5}\nonumber\\
&+\left(\frac{-x-y+1}{x^2 (x+y-z+1)}+\frac{1}{2 x (x+y-z)}\right)f^{\textup{t1}\prime}_{6}
+\left(\frac{1}{2 x (x+y-z)}\right)f^{\textup{t1}\prime}_{7}\nonumber\\
&+\left(\frac{2 x+y-2 z}{2 x (x-z) (x+y-z)}-\frac{3 (x+y-1)}{2 x^2 (x+y-z+1)}\right)f^{\textup{t1}\prime}_{8}
+\left(\frac{-13 x^2+x (19 z-12 y)+6 z (y-z)}{2 x^2 (x-z) (x+y-z)}\right)f^{\textup{t1}\prime}_{9}\nonumber\\
&+\left(\frac{-x-y+1}{4 x^2 (x+y-z+1)}+\frac{x-y-z}{4 x (x-z) (x+y-z)}\right)f^{\textup{t1}\prime}_{10}
+\left(-\frac{y}{2 x (x-z) (x+y-z)}\right)f^{\textup{t1}\prime}_{11}\nonumber\\
&+\left(-\frac{y}{2 x (x-z) (x+y-z)}\right)f^{\textup{t1}\prime}_{12}
+\left(\frac{12 x-6 z}{4 x^2 (x-z)}\right)f^{\textup{t1}\prime}_{14}
+\left(-\frac{2 (2 x+(y-1) z)}{x^2 (x-z) (x+y-z+1)}\right)f^{\textup{t1}\prime}_{15}\nonumber\\
&+\left(\frac{6 (x+y-1)}{x^2 (x+y-z+1)}\right)f^{\textup{t1}\prime}_{16}
+\left(\frac{x+4 y-z}{2 x (x-z) (x+y-z)}\right)f^{\textup{t1}\prime}_{17}\nonumber\\
&+\left(\frac{3 (x+y-1)}{2 x^2 (x+y-z+1)}+\frac{-x+y+z}{2 x (x-z) (x+y-z)}\right)f^{\textup{t1}\prime}_{19}\nonumber\\
&+\left(\frac{-x-y+1}{2 x^2 (x+y-z+1)}+\frac{x-y-z}{2 x (x-z) (x+y-z)}\right)f^{\textup{t1}\prime}_{20}
+\left(\frac{1}{2 x (x-z)}-\frac{3 (x+y-1)}{2 x^2 (x+y-z+1)}\right)f^{\textup{t1}\prime}_{21}\nonumber\\
&+\left(\frac{y}{x (x-z) (x+y-z)}\right)f^{\textup{t1}\prime}_{22}
+\left(\frac{1}{x (x-z)}\right)f^{\textup{t1}\prime}_{23}
+\left(\frac{1}{x (x-z)}\right)f^{\textup{t1}\prime}_{24}
+\left(\frac{x+y-1}{x^2 (x+y-z+1)}\right)f^{\textup{t1}\prime}_{25}\nonumber\\
&+\left(\frac{x+y-1}{x^2 (x+y-z+1)}\right)f^{\textup{t1}\prime}_{26}
+\left(\frac{1}{x (x-z)}\right)f^{\textup{t1}\prime}_{27}
+\left(\frac{-x-y+1}{x^2 (x+y-z+1)}\right)f^{\textup{t1}\prime}_{29}
+\left(-\frac{1}{x (x-z)}\right)f^{\textup{t1}\prime}_{30}.
\end{align}
\normalsize

\paragraph*{Topology 2:}

\footnotesize
\begin{align}
f^\textup{t2}_{8}=&
\left(\frac{5\epsilon^2 (\epsilon (x (2 z-1)+(5-2 z) z)+z (-x+z-2))}{(2\epsilon-1) (3\epsilon-2) (3\epsilon-1) z}\right)f^{\textup{t2}\prime}_{1}
+\left(\frac{10\epsilon^3 x (z-1)}{(2\epsilon-1) (3\epsilon-2) (3\epsilon-1) z}\right)f^{\textup{t2}\prime}_{6}\nonumber\\
&+\left(\frac{5\epsilon^2 (z-1) ((z-1) z-x (z+1))}{(3\epsilon-2) (3\epsilon-1) z^2}\right)f^{\textup{t2}\prime}_{7}\nonumber\\
&+\left(\frac{5\epsilon^2 (x-z+1) (\epsilon (x (2 z-1)-2 (z-1) z)+z (-x+z-1))}{(2\epsilon-1) (3\epsilon-2) (3\epsilon-1) z (x-z)}\right)f^{\textup{t2}\prime}_{8},
\end{align}
\normalsize

\footnotesize
\begin{align}
f^\textup{t2}_{21}=&
\left(\frac{5\epsilon^2 \left(\epsilon^2 \left(2 z^3+51 z^2+12 z-1\right)-\epsilon \left(2 z^3+57 z^2+6 z-1\right)+12 z^2\right) (x+y)}{3 (\epsilon-1) (2\epsilon-1) (3\epsilon-2) (3\epsilon-1) z^2}\right)f^{\textup{t2}\prime}_{20}\nonumber\\
&+\left(-\frac{5\epsilon^3 (z-1) \left(\epsilon \left(2 z^2+7 z-1\right)-2 z^2-5 z+1\right) (x+y)}{3 (\epsilon-1) (2\epsilon-1) (3\epsilon-2) (3\epsilon-1) z^2}\right)f^{\textup{t2}\prime}_{21},
\end{align}
\normalsize

\footnotesize
\begin{align}
f^\textup{t2}_{34}=&
\left(-\frac{40 (x+y) (x+y+1)}{3 x^2 (z-1) (x+y-z+1)}\right)f^{\textup{t2}\prime}_{7}
+\left(-\frac{40 (x+y) (x+y+1)}{3 x^2 (z-1) (x+y-z+1)}\right)f^{\textup{t2}\prime}_{8}\nonumber\\
&+\left(\frac{5 (x+y) (x+y+1)}{x^2 (z-1) (x+y-z+1)}\right)f^{\textup{t2}\prime}_{11}
+\left(\frac{5 (x+y+1) (5 x+5 y-8 z+8)}{x^2 (z-1) (x+y-z+1)}\right)f^{\textup{t2}\prime}_{12}\nonumber\\
&+\left(\frac{35 (x+y) (x+y+1)}{22 x^2 (z-1) (x+y-z+1)}\right)f^{\textup{t2}\prime}_{13}
+\left(\frac{10 (y+1)}{x^2 (y-1)}-\frac{5 (x+y+1)}{2 x^2 (z-1)}\right)f^{\textup{t2}\prime}_{14}\nonumber\\
&+\left(-\frac{7 (x+y+1)}{2 x^2 (x+y-z+1)}\right)f^{\textup{t2}\prime}_{23}
+\left(\frac{2 (x+y+1)}{x^2 (x+y-z+1)}+\frac{71 (y+1)}{2 x^2 (y-1)}\right)f^{\textup{t2}\prime}_{24}\nonumber\\
&+\left(\frac{2 (x (5 y+4)+(y+1) (5 y-5 z+4))}{x^2 (x+y-z+1)}\right)f^{\textup{t2}\prime}_{29}
+\left(-\frac{20 (x+y) (x+y+1)}{x^2 (z-1) (x+y-z+1)}\right)f^{\textup{t2}\prime}_{30}\nonumber\\
&+\left(-\frac{5 z}{x^2 (x+y-z+1)}\right)f^{\textup{t2}\prime}_{32}
+\left(-\frac{10 (x+y+1)}{x^2 (x+y-z+1)}\right)f^{\textup{t2}\prime}_{33}
+\left(-\frac{10 (y+1)}{x^2 (y-1)}\right)f^{\textup{t2}\prime}_{35}.
\end{align}
\normalsize
The primed integrals denote integrals of the canonical basis. In the resulting $\epsilon$-form of the differential equations, the following sets of letters have non-vanishing coefficient matrices

\begin{eqnarray}
\mathcal{A}^\textup{t1}&=&\big\{x,\,y,\,x+y,\,x-z,\,y-z,\,x+y-z,\,1+x+y-z,\,-1+z,\nonumber \\ 
&&z,\,-1-x+z,\,y(-1+z)+(1+x-z)z\big\},
\end{eqnarray}

\begin{eqnarray}
\mathcal{A}^\textup{t2}&=&\big\{x,\,-1+y,\,y,\,x+y,\,x-z,\,1+x-z,\,y-z,\,x+y-z,\nonumber \\ 
&&1+x+y-z,\,-1+z,\,z,\,x+y(1-z),\,x(-1+y)+y(y-z),\nonumber \\ 
&&y(-1+z)+(1+x-z)z\big\}.
\end{eqnarray}

\subsection{Vector boson pair production}
The second set of examples has been used in the computation of the NNLO QCD corrections to the production of two massive vector bosons \cite{Cascioli:2014yka, Gehrmann:2014fva, Grazzini:2016swo}. These integral topologies have been considered in \cite{Gehrmann:2013cxs, Henn:2014lfa, Gehrmann:2014bfa, Caola:2014lpa, Papadopoulos:2014hla, Gehrmann:2015ora} and are given by
\paragraph*{Topology 1:}
\begin{equation}
\begin{array}{lll}
P_1=l_1^2, & P_4=(l_2-p_3-p_4)^2, & P_7=(l_2-p_1)^2, \\
P_2=(l_1-p_3-p_4)^2, & P_5=(l_1-p_3)^2, &  P_8=(l_2-p_3)^2, \\
P_3=l_2^2, &P_6=(l_1-l_2)^2, & P_9=(l_1-p_1)^2. \\  
\end{array}
\end{equation}
\paragraph*{Topology 2:}
\begin{equation}
\begin{array}{lll}
P_1=l_1^2, & P_4=(l_2+p_1-p_3)^2, & P_7=(l_2+p_4)^2, \\
P_2=(l_1+p_1-p_3)^2, & P_5=(l_1-p_3)^2, &  P_8=(l_2-p_3)^2, \\
P_3=l_2^2, &P_6=(l_1-l_2)^2, & P_9=(l_1+p_4)^2. \\  
\end{array}
\end{equation}
As before, the momenta $p_1$ and $p_2$ are incoming and $p_3$ and $p_4$ are outgoing. The kinematics of both topologies are given by
\begin{equation}
p_1^2=0,\quad p_2^2=0,\quad p_3^2=m_3^2,\quad p_4^2=m_4^2,
\end{equation}
\begin{equation}
p_1+p_2=p_3+p_4,
\end{equation}
\begin{equation}
s\defeq(p_1+p_2)^2,\quad t\defeq(p_1-p_3)^2.
\end{equation}
\begin{figure}[ht]
\begin{center}
\fcolorbox{white}{white}{
  \begin{picture}(330,120) (-5,-70)
   \SetScale{0.5}
   \unitlength=0.5pt
    \SetWidth{3.0}
    \SetColor{Black}
    \Line(96,51)(454,51)
    \Photon(454,-83)(544,-83){7.5}{7}
    \Photon(454,51)(544,51){7.5}{7}
    \Line(454,-83)(454,51)
    \Line(96,-83)(455,-83)
    \Text(564,-93)[lb]{\large{\Black{$p_4$}}}
    \Text(564,41)[lb]{\large{\Black{$p_3$}}}
    \Text(55,41)[lb]{\large{\Black{$p_1$}}}
    \Text(55,-93)[lb]{\large{\Black{$p_2$}}}
    \Line(186,51)(186,-83)
    \Line(320,51)(320,-83)
  \end{picture}
}
\end{center}
\caption{Two loop graph of topology 1.}
\end{figure}
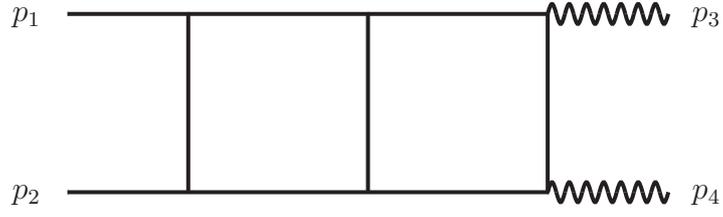
\begin{figure}[ht]
\begin{center}
\fcolorbox{white}{white}{
  \begin{picture}(330,120) (-5,-70)
   \SetScale{0.5}
   \unitlength=0.5pt
    \SetWidth{3.0}
    \SetColor{Black}
    \Line(184,51)(454,51)
	\Photon(96,51)(186,51){7.5}{7}
    \Photon(454,51)(544,51){7.5}{7}
    \Line(454,-83)(454,51)
    \Line(96,-83)(544,-83)
    \Text(564,-93)[lb]{\large{\Black{$p_1$}}}
    \Text(564,41)[lb]{\large{\Black{$p_3$}}}
    \Text(55,41)[lb]{\large{\Black{$p_4$}}}
    \Text(55,-93)[lb]{\large{\Black{$p_2$}}}
    \Line(186,51)(186,-83)
    \Line(320,51)(320,-83)
  \end{picture}
}
\end{center}
\caption{Two loop graph of topology 2.}
\end{figure}
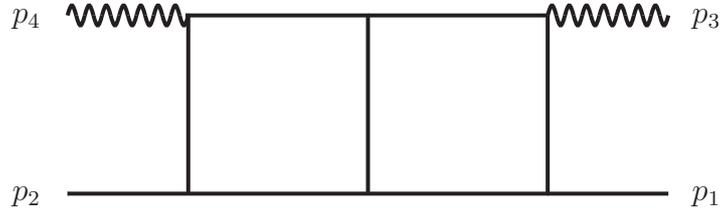
Topology 1 has a basis of 31 master integrals:
\begin{equation}
\label{OldBasisVVT1}
\begin{array}{lll}
\vec{g}^{\,\textup{t1}}(\epsilon,s,t,m_3,m_4)\,\,\,=&\big(I_\textup{t1}(1,1,1,1,0,0,0,0,0),&I_\textup{t1}(1,0,1,1,1,0,0,0,0),\\
&I_\textup{t1}(0,1,1,1,1,0,0,0,0),&I_\textup{t1}(1,1,1,1,1,0,0,0,0),\\
&I_\textup{t1}(0,1,1,0,0,1,0,0,0),&I_\textup{t1}(0,0,1,0,1,1,0,0,0),\\
&I_\textup{t1}(-1,1,1,0,1,1,0,0,0),&I_\textup{t1}(0,1,1,0,1,1,0,0,0),\\
&I_\textup{t1}(0,0,0,1,1,1,0,0,0),&I_\textup{t1}(1,-1,0,1,1,1,0,0,0),\\
&I_\textup{t1}(1,0,0,1,1,1,0,0,0),&I_\textup{t1}(-1,0,1,1,1,1,0,0,0),\\
&I_\textup{t1}(0,0,1,1,1,1,0,0,0),&I_\textup{t1}(1,0,1,1,1,1,0,0,0),\\
&I_\textup{t1}(0,1,1,1,1,1,0,0,0),&I_\textup{t1}(1,1,0,0,0,1,1,0,0),\\
&I_\textup{t1}(0,0,0,0,1,1,1,0,0),&I_\textup{t1}(1,0,0,0,1,1,1,0,0),\\
&I_\textup{t1}(0,1,0,0,1,1,1,0,0),&I_\textup{t1}(1,1,-1,0,1,1,1,0,0),\\
&I_\textup{t1}(1,1,0,0,1,1,1,0,0),&I_\textup{t1}(-1,1,1,0,1,1,1,0,0),\\
&I_\textup{t1}(0,1,1,0,1,1,1,0,0),&I_\textup{t1}(1,-1,0,1,1,1,1,0,0),\\
&I_\textup{t1}(1,0,0,1,1,1,1,0,0),&I_\textup{t1}(0,0,1,1,1,1,1,0,0),\\
&I_\textup{t1}(1,0,1,1,1,1,1,0,0),&I_\textup{t1}(0,1,1,1,1,1,1,0,0),\\
&I_\textup{t1}(1,1,1,1,1,1,1,-1,0),&I_\textup{t1}(1,1,1,1,1,1,1,0,-1),\\
&I_\textup{t1}(1,1,1,1,1,1,1,0,0)\big).&
\end{array}
\end{equation}
Topology 2 has a basis of 29 master integrals:
\begin{equation}
\label{OldBasisVVT2}
\begin{array}{lll}
\vec{g}^{\,\textup{t2}}(\epsilon,s,t,m_3,m_4)\,\,\,=&\big(I_\textup{t2}(1,1,1,1,0,0,0,0,0),&I_\textup{t2}(1,0,1,1,1,0,0,0,0),\\
&I_\textup{t2}(0,1,1,0,0,1,0,0,0),&I_\textup{t2}(0,0,1,0,1,1,0,0,0),\\
&I_\textup{t2}(1,0,0,1,1,1,0,0,0),&I_\textup{t2}(0,0,1,1,1,1,0,0,0),\\
&I_\textup{t2}(1,0,1,1,1,1,0,0,0),&I_\textup{t2}(1,1,1,0,0,0,1,0,0),\\
&I_\textup{t2}(1,0,1,0,1,0,1,0,0),&I_\textup{t2}(1,0,0,0,0,1,1,0,0),\\
&I_\textup{t2}(1,1,0,0,0,1,1,0,0),&I_\textup{t2}(0,1,1,0,0,1,1,0,0),\\
&I_\textup{t2}(1,1,1,0,0,1,1,0,0),&I_\textup{t2}(0,0,0,0,1,1,1,0,0),\\
&I_\textup{t2}(1,-1,0,0,1,1,1,0,0),&I_\textup{t2}(1,0,0,0,1,1,1,0,0),\\
&I_\textup{t2}(1,1,0,0,1,1,1,0,0),&I_\textup{t2}(-1,0,1,0,1,1,1,0,0),\\
&I_\textup{t2}(0,0,1,0,1,1,1,0,0),&I_\textup{t2}(1,0,1,0,1,1,1,0,0),\\
&I_\textup{t2}(-1,1,1,0,1,1,1,0,0),&I_\textup{t2}(0,1,1,0,1,1,1,0,0),\\
&I_\textup{t2}(1,1,1,0,1,1,1,0,0),&I_\textup{t2}(1,-1,0,1,1,1,1,0,0),\\
&I_\textup{t2}(1,0,0,1,1,1,1,0,0),&I_\textup{t2}(0,0,1,1,1,1,1,0,0),\\
&I_\textup{t2}(1,0,1,1,1,1,1,0,0),&I_\textup{t2}(1,1,1,1,1,1,1,-1,0),\\
&I_\textup{t2}(1,1,1,1,1,1,1,0,0)\big).&
\end{array}
\end{equation}
The mass-dimension is factored out of the master integrals as follows
\begin{equation}
f_i(\epsilon,x,y,z)=(m_3)^{-\textup{dim}(g_i)}g_i(\epsilon,s,t,m_3,m_4).
\end{equation}
The set of dimensionless parameters is taken to be the same as in \cite{Henn:2014lfa}
\begin{equation}
(1+x)(1+xy)=\frac{s}{m_3^2},\quad -xz=\frac{t}{m_3^2},\quad x^2y=\frac{m_4^2}{m_3^2}.
\end{equation}
The transformations to canonical bases have been computed with an implementation of the algorithm from section \ref{sec:Algorithm} in \texttt{Mathematica}. In the following, the decomposition of three integrals of each topology in terms of the canonical bases is shown. The full transformations are provided in ancillary files.
\paragraph*{Topology 1:}

\footnotesize
\begin{align}
f^\textup{t1}_{8}=&
\left(\frac{2\epsilon^2 (x+1)}{(2\epsilon-1) (3\epsilon-1) x (y-1)}\right)f^{\textup{t1}\prime}_{5}
+\left(-\frac{2\epsilon^2 (x y+1)}{(2\epsilon-1) (3\epsilon-1) x (y-1)}\right)f^{\textup{t1}\prime}_{6}\nonumber\\
&+\left(\frac{2\epsilon^2}{(2\epsilon-1) (3\epsilon-1)}\right)f^{\textup{t1}\prime}_{7}
+\left(-\frac{2\epsilon^2 (x+1)}{(2\epsilon-1) (3\epsilon-1) x (y-1)}\right)f^{\textup{t1}\prime}_{8},
\end{align}
\normalsize

\footnotesize
\begin{align}
f^\textup{t1}_{10}=&
\left(\frac{2\epsilon^2 \left(\epsilon \left(5 x^2 y+3 x (y+1)+1\right)-x (2 x y+y+1)\right)}{(2\epsilon-1) (3\epsilon-2) (3\epsilon-1)}\right)f^{\textup{t1}\prime}_{5}\nonumber\\
&+\left(\frac{2\epsilon^3 x^2 y}{(2\epsilon-1) (3\epsilon-2) (3\epsilon-1)}\right)f^{\textup{t1}\prime}_{9}
+\left(\frac{\epsilon^2 x (2 x y+y+1)}{(3\epsilon-2) (3\epsilon-1)}\right)f^{\textup{t1}\prime}_{10}\nonumber\\
&+\left(-\frac{\epsilon^2 x (x y+1) (\epsilon ((x-1) y+2)-x y-1)}{(2\epsilon-1) (3\epsilon-2) (3\epsilon-1) (y-1)}\right)f^{\textup{t1}\prime}_{11},
\end{align}
\normalsize

\footnotesize
\begin{align}
f^\textup{t1}_{31}=&
\left(-\frac{1}{2 x (x+1)^2 z (x y+1)^2}\right)f^{\textup{t1}\prime}_{9}
+\left(-\frac{2}{x (x+1)^2 z (x y+1)^2}\right)f^{\textup{t1}\prime}_{11}\nonumber\\
&+\left(\frac{2}{3 x (x+1)^2 z (x y+1)^2}\right)f^{\textup{t1}\prime}_{19}
+\left(\frac{2}{x (x+1)^2 z (x y+1)^2}\right)f^{\textup{t1}\prime}_{23}\nonumber\\
&+\left(\frac{2}{x (x+1)^2 z (x y+1)^2}\right)f^{\textup{t1}\prime}_{24}
+\left(\frac{10}{x (x+1)^2 z (x y+1)^2}\right)f^{\textup{t1}\prime}_{25}\nonumber\\
&+\left(-\frac{2}{x (x+1)^2 z (x y+1)^2}\right)f^{\textup{t1}\prime}_{29}.
\end{align}
\normalsize

\paragraph*{Topology 2:}

\footnotesize
\begin{align}
f^\textup{t2}_{16}=&
\left(\frac{\epsilon^2}{(2\epsilon-1) (3\epsilon-1)}\right)f^{\textup{t2}\prime}_{14}
+\left(\frac{\epsilon (\epsilon x y+\epsilon)}{(2\epsilon-1) (3\epsilon-1) (y-1)}\right)f^{\textup{t2}\prime}_{15}\nonumber\\
&+\left(-\frac{\epsilon^2}{(2\epsilon-1) (3\epsilon-1)}\right)f^{\textup{t2}\prime}_{16},
\end{align}
\normalsize

\footnotesize
\begin{align}
f^\textup{t2}_{18}=&
\left(\frac{\epsilon^2 \left(\epsilon \left(x^2 y (2 y-1)-x y-2\right)-x^2 y^2+1\right)}{2 (2\epsilon-1) (3\epsilon-2) (3\epsilon-1) x (y-1)}\right)f^{\textup{t2}\prime}_{4}
+\left(\frac{\epsilon^2 (\epsilon (2 x (y+1)+1)-x (y+1))}{2 (2\epsilon-1) (3\epsilon-2) (3\epsilon-1)}\right)f^{\textup{t2}\prime}_{18}\nonumber\\
&+\left(\frac{\epsilon^2 (x+1) (\epsilon (x (y-2)+2)+x-1)}{2 (2\epsilon-1) (3\epsilon-2) (3\epsilon-1) x (y-1)}\right)f^{\textup{t2}\prime}_{19},
\end{align}
\normalsize

\footnotesize
\begin{align}
f^\textup{t2}_{29}=&
\left(\frac{1}{x^2 (x+1) z^2 (x y+1)}\right)f^{\textup{t2}\prime}_{9}
+\left(\frac{1}{x^2 (x+1) z^2 (x y+1)}\right)f^{\textup{t2}\prime}_{18}\nonumber\\
&+\left(-\frac{1}{x^2 (x+1) z^2 (x y+1)}\right)f^{\textup{t2}\prime}_{20}
+\left(-\frac{2}{x^2 (x+1) z^2 (x y+1)}\right)f^{\textup{t2}\prime}_{23}\nonumber\\
&+\left(-\frac{2}{x^2 (x+1) z^2 (x y+1)}\right)f^{\textup{t2}\prime}_{29}.
\end{align}
\normalsize
In the canonical bases, the resulting differential equations are in $\epsilon$-form and the following sets of letters have non-vanishing coefficients matrices
\begin{eqnarray}
\mathcal{A}^\textup{t1}&=&\big\{x,\,1+x,\,1-y,\,y,\,1+xy,\,1+x(1+y-z),\,1-z,\nonumber\\
&&1+(1+x)y-z,\,z,\,z-y,\,z+xy,\,1+xz\big\},
\end{eqnarray}
\begin{eqnarray}
\mathcal{A}^\textup{t2}&=&\big\{x,\,1+x,\,1-y,\,y,\,1+xy,\,1+x(1+y-z),\,1-z,\nonumber\\
&&1+(1+x)y-z,\,z,\,z-y,\,z+xy,\,1+xz,\nonumber\\
&&z-y+yz+xyz,\,z-xy+xz+xyz\big\}.
\end{eqnarray}

\section{Conclusion}
\label{sec:Conclusion}
Assuming the existence of a rational transformation that transforms a differential equation of master integrals to an $\epsilon$-form, the algorithm presented here can be used to compute such a transformation. It is applicable to differential equations involving multiple scales and allows for a rational dependence of the differential equation on the dimensional regulator and thus extends previous approaches. It has been shown that the transformation can be obtained as the solution of finitely many differential equations. These are solved with an ansatz that is given by a linear combination of rational functions in Leinartas form. Any multivariate rational function can be expressed as a linear combination of functions of this type. After choosing a sufficiently large set of these functions for the ansatz, a transformation can be constructed by solving polynomial equations in the parameters of the ansatz. As already suggested in previous approaches, it is beneficial to make use of the block-triangular form of the differential equation by computing the transformation in a recursion over subsectors. This strategy has been incorporated into the presented algorithm as well.

The algorithm has been implemented in \texttt{Mathematica}, which will be the topic of a further publication \cite{Meyer:0000abc}. The power of the algorithm has been demonstrated by its application to non-trivial integral topologies, some of which were previously unknown. With its broad scope of application, the presented algorithm may prove particularly useful to facilitate multi-loop calculations that involve multiple scales.

\acknowledgments
The author would like to thank P. Uwer for useful discussions and V. Smirnov and P. Uwer for comments on the manuscript. The Feynman diagrams in this paper have been drawn with \texttt{JAXODRAW} \cite{Binosi:2008ig, Vermaseren:1994je}. This research was supported by the DFG via the Research Training Group 1504.

\appendix

\section{Polynomial rings} \label{App:PolyAlgebra}
For convenience, this appendix reproduces some standard definitions and results about polynomial algebra, which are used in section \ref{sec:Leinartas}. For a more detailed exposition the reader is referred to \cite{Cox:2007:IVA:1204670}.
\begin{defn}[Ideal]
A subset $I\subseteq K[X]$ is called an ideal if the following conditions are satisfied
\begin{enumerate}
\item $0\in I$.
\item If $f,g\in I$, then $f+g\in I$.
\item If $f\in I$ and $h\in K[X]$, then $hf\in I$.
\end{enumerate}
\end{defn}
\begin{defn}[Ideal generated by a set of polynomials]
Let $\{f_1,\dots,f_m\}\subset K[X]$ be a set of polynomials. Then 
\begin{equation*}
\langle f_1,\dots,f_m\rangle=\left\{\sum_{i=1}^mh_if_i \quad\big|\quad h_1,\dots h_m\in K[X]\right\}
\end{equation*}
is an ideal, which is called the ideal generated by $\{f_1,\dots,f_m\}$.
\end{defn}
\begin{defn}[Irreducible polynomial]
A polynomial $f\in K[X]$ is called irreducible over $K$, if f is non-constant and is not the product of two non-constant polynomials in $K[X]$.
\end{defn}
\begin{thm}[Factorization]
Every non-constant $f\in K[X]$ can be written as a product $f=f_1^{e_1}\dots f_m^{e_m}$ of irreducible polynomials over $K$. This factorization is unique up to multiplication with constant factors and reordering of the irreducible factors $f_i$. 
\end{thm}
\begin{thm}[Weak Nullstellensatz]
Let $I\subseteq K[X]$ be an ideal that satisfies 
\begin{equation*}
V(I)=\cap_{f\in I}V(f)=\emptyset,
\end{equation*}
then $I=K[X]$.
\end{thm}

\begin{defn}[Monomial ordering]
\label{def:monOrd}
A monomial ordering on $K[X]$ is a relation $>$ on the set of monomials $x^\alpha$, $\alpha\in\mathbb{Z}^d_{\geq 0}$ which satisfies:
\begin{enumerate}
\item $>$ is a total ordering on $\mathbb{Z}^d_{\geq 0}$.
\item \label{def:monOrd:2} If $\alpha>\beta$ and $\gamma\in\mathbb{Z}^d_{\geq 0}$, then $\alpha+\gamma>\beta+\gamma$.
\item \label{def:monOrd:3} For all $A\subseteq\mathbb{Z}^d_{\geq 0}$ there exists an $\alpha\in A$ such that $\beta>\alpha$ for all $\beta\neq\alpha$ in A.
\end{enumerate}
\end{defn}

While the considerations in subsection \ref{sec:Leinartas} are agnostic about the monomial ordering, in practice the \emph{lexicographic ordering} has proven to be a good choice.
\begin{defn}[Lexicographic ordering]
For $\alpha=(\alpha_1,\dots,\alpha_d)$ and $\beta=(\beta_1,\dots,\beta_d)$ in $\mathbb{Z}^d_{\geq 0}$ it is said that $\alpha>_\textup{lex}\beta$, if the leftmost nonzero entry of $\alpha-\beta\in\mathbb{Z}^d$ is positive.
\end{defn}
Note that different orderings of the variables give rise to different lexicographic orderings.
\begin{defn}[]
Let $f=\sum_\alpha a_\alpha x^\alpha$ be a nonzero polynomial in $K[X]$ and $\alpha\in\mathbb{Z}^d_{\geq 0}$ and let $>$ be a monomial order. 
\begin{enumerate}
\item The multidegree of $f$ is
\begin{equation*}
\textup{multideg}(f)=\textup{max}\left\{\alpha\in\mathbb{Z}^d_{\geq 0}\,|\,a_\alpha\neq 0\right\}
\end{equation*}
the maximum is taken with respect to the monomial order $>$.
\item The leading coefficient of $f$ is
\begin{equation*}
\textup{LC}(f)=a_{\textup{multideg}(f)}\in K.
\end{equation*}
\item The leading monomial of $f$ is
\begin{equation*}
\textup{LM}(f)=x^{\textup{multideg}(f)}.
\end{equation*}
\item The leading term of $f$ is
\begin{equation*}
\textup{LT}(f)=\textup{LC}(f)\cdot\textup{LM}(f).
\end{equation*}
\end{enumerate}
\end{defn}
\begin{lemma}
\label{lem:multidegProperties}
Let $f,g\in K[X]$ be nonzero polynomials. Then 
\begin{equation*}
\textup{multideg}(fg)=\textup{multideg}(f)+\textup{multideg}(g).
\end{equation*}
\end{lemma}

\begin{lemma}
\label{lem:lowerboundmonoordering}
Let $>$ be a relation on $\mathbb{Z}^d_{\geq 0}$ satisfying: 
\begin{enumerate}
\item $>$ is a total ordering on $\mathbb{Z}^d_{\geq 0}$.
\item If $\alpha>\beta$ and $\gamma\in\mathbb{Z}^d_{\geq 0}$, then $\alpha+\gamma>\beta+\gamma$.
\end{enumerate}
Then $\alpha\geq 0$ for all $\alpha\in\mathbb{Z}^d_{\geq 0}$ if and only if for all $A\subseteq\mathbb{Z}^d_{\geq 0}$ there exists an $\alpha\in A$ such that $\beta>\alpha$ for all $\beta\neq\alpha$ in A.
\end{lemma}
This lemma implies that $\alpha\geq 0$ holds for any monomial ordering and for all $\alpha\in\mathbb{Z}^d_{\geq 0}$.

\begin{defn}[Set of leading terms]
Fix a monomial ordering on $K[X]$ and let $I\subseteq K[X]$ be an ideal other than $\{0\}$, then $\textup{LT}(I)$ denotes the set of leading terms of nonzero elements of $I$, i.e.
\begin{equation*}
\textup{LT}(I)=\left\{cx^\alpha\,\,|\,\,\exists f \in I\setminus \{0\}\,\, \textup{with}\,\, \textup{LT}(f)=cx^\alpha\right\}.
\end{equation*}
\end{defn}

\begin{defn}[Gr\"obner basis]
Fix a monomial ordering on $K[X]$. A finite subset $G=\{g_1,\dots,g_t\}$ of an ideal $I\subseteq K[X]$ other than $\{0\}$ is said to be a \emph{Gr\"obner basis} if
\begin{equation*}
\langle\textup{LT}(g_1),\dots,\textup{LT}(g_t)\rangle=\langle\textup{LT}(I)\rangle.
\end{equation*}
\end{defn}

\bibliographystyle{JHEP}
\bibliography{../../literature.bib}


\end{document}